\documentclass[twocolumn,superscriptaddress,amsmath,amssymb,aps,pra]{revtex4-1}

\usepackage{graphicx}
\usepackage{bm}

\usepackage{color}

\usepackage{graphicx}
\usepackage{comment}
\usepackage{color}
\usepackage[colorlinks,urlcolor=blue,citecolor=blue,linkcolor=blue]{hyperref}
\usepackage{cleveref}
\usepackage{physics}
\usepackage{mathtools}
\usepackage{mathrsfs}
\usepackage{booktabs}

\newcommand{\kv}{\vectorbold{k}}
\newcommand{\Kv}{\vectorbold{K}}
\newcommand{\kvt}{\vectorbold{K}_{3\rm{D}}}
\newcommand{\qv}{\vectorbold{q}}
\newcommand{\Qv}{\vectorbold{Q}}
\newcommand{\qvt}{\vectorbold{Q}_{3\rm{D}}}
\newcommand{\p}{\vectorbold{p}}
\newcommand{\sv}{\vectorbold{s}}
\newcommand{\xv}{\vectorbold{x}}
\newcommand{\zerov}{\vectorbold{0}}

\renewcommand{\pv}{\vectorbold{p}}
\newcommand{\nuv}{\boldsymbol{\nu}}

\newcommand{\ch}{\hat{c}}
\newcommand{\chd}{\hat{c}^\dagger}

\renewcommand{\dh}{\hat{d}}
\newcommand{\dhd}{\hat{d}^\dagger}

\newcommand{\Dfreeh}{\hat{D}^{(0)}}
\newcommand{\Gfreeh}{\hat{G}^{(0)}}
\newcommand{\Hh}{\hat{H}}
\newcommand{\Th}{\hat{T}}
\newcommand{\Dh}{\hat{D}}
\newcommand{\Pih}{\hat{\Pi}}
\newcommand{\Sv}{\vectorbold{S}}
\newcommand{\Nv}{\vectorbold{N}}
\newcommand{\nv}{\vectorbold{n}}

\newcommand{\new}[1]{{\color{black}#1}}

\begin{document}
\title{Microscopic theory of the Hubbard interaction in low-dimensional optical lattices}

\author{Haydn S. Adlong}
\affiliation{School of Physics and Astronomy, Monash University, Victoria 3800, Australia}
\affiliation{ARC Centre of Excellence in Future Low-Energy Electronics Technologies, Monash University, Victoria 3800, Australia}
\affiliation{Institute for Quantum Electronics, ETH Z\"{u}rich, CH-8093 Z\"{u}rich, Switzerland}
\affiliation{Institute for Theoretical Physics, ETH Z\"{u}rich, CH-8093 Z\"{u}rich, Switzerland}

\author{Jesper Levinsen}
\affiliation{School of Physics and Astronomy, Monash University, Victoria 3800, Australia}
\affiliation{ARC Centre of Excellence in Future Low-Energy Electronics Technologies, Monash University, Victoria 3800, Australia}

\author{Meera M. Parish}
\affiliation{School of Physics and Astronomy, Monash University, Victoria 3800, Australia}
\affiliation{ARC Centre of Excellence in Future Low-Energy Electronics Technologies, Monash University, Victoria 3800, Australia}

\begin{abstract}
    The Hubbard model is a paradigmatic model of strongly correlated quantum matter, thus making it desirable to investigate with quantum simulators such as ultracold atomic gases. Here, we consider the problem of two atoms interacting in a quasi-one- or quasi-two-dimensional optical lattice, 
    geometries which are routinely realized in quantum-gas-microscope experiments.
    We perform an exact calculation of the low-energy scattering amplitude which accounts for the effects of the transverse 
    confinement as well as all higher Bloch bands. 
    This goes beyond standard perturbative treatments and allows us to precisely determine the effective Hubbard on-site interaction for arbitrary $s$-wave scattering length (see source code available at~\cite{SourceCode}). 
    In particular, we find that the Hubbard on-site interaction displays lattice-induced resonances for scattering lengths on the order of the lattice spacing, which are well within reach of current experiments. 
    Furthermore, we show that our results are in excellent agreement with spectroscopic measurements of the Hubbard interaction for a quasi-two-dimensional square optical lattice in a quantum gas microscope.  
    Our formalism is very general and may be extended to multi-band models and other atom-like scenarios in lattice geometries, such as exciton-exciton and exciton-electron scattering in moir\'e superlattices.
\end{abstract}

\date{\today}

\maketitle

\section{Introduction}

Ultracold atoms in optical lattices are a promising platform for the simulation of strongly correlated matter beyond the reach of conventional computation.
The power of these experiments lies in their remarkable controllability and precision measurement techniques~\cite{blochQuantumSimulationsUltracold2012,grossQuantumSimulationsUltracold2017,schaferToolsQuantumSimulation2020}, which enable the realization of models ranging from solid-state physics to high energy physics and astrophysics, thus making them ideal quantum analog simulators~\cite{lewensteinUltracoldAtomicGases2007, blochManybodyPhysicsUltracold2008}. 
In particular, for sufficiently deep optical lattices, cold-atom experiments can simulate the Hubbard model~\cite{jakschColdAtomHubbard2005}, one of the simplest models that captures important many-body phenomena such as magnetism and superconductivity. 
Notable early achievements include the realization of the Mott insulating state in both Bose and Fermi Hubbard systems~\cite{greinerQuantumPhaseTransition2002,jordensMottInsulatorFermionic2008,schneiderMetallicInsulatingPhases2008}, and the observation of antiferromagnetic correlations in the fermionic case~\cite{greifShortRangeQuantumMagnetism2013,hartObservationAntiferromagneticCorrelations2015}.

The advent of quantum gas microscopes~\cite{bakrQuantumGasMicroscope2009} has further advanced the capabilities of 
optical lattices as Hubbard-model simulators, since they allow a quantum many-body state such as a Mott insulator to be imaged at the single-atom level~\cite{shersonSingleatomresolvedFluorescenceImaging2010,bakrProbingSuperfluidMott2010,weitenbergSinglespinAddressingAtomic2011,cheukObservation2DFermionic2016,greifSiteresolvedImagingFermionic2016}.
Thus, quantum gas microscopes offer unprecedented access to the atoms' spatial distribution and correlations in the Hubbard regime. This includes antiferromagnetic~\cite{bollSpinDensityresolvedMicroscopy2016,parsonsSiteresolvedMeasurementSpincorrelation2016, cheukObservationSpatialCharge2016} and ferromagnetic~\cite{Lebrat2024,Prichard2024} correlations,
as well as less visible correlated phenomena such as non-local string order from correlated particle-hole pairs~\cite{endresObservationCorrelatedParticleHole2011,hilkerRevealingHiddenAntiferromagnetic2017}, and entangled many-body localized states~\cite{kaufmanQuantumThermalizationEntanglement2016,lukinProbingEntanglementManybody2019,rispoliQuantumCriticalBehaviour2019} 
(see Refs.~\cite{grossQuantumGasMicroscopy2021,schaferToolsQuantumSimulation2020} for recent reviews).

Despite the significant experimental progress, there is still a dearth of microscopic theories that can accurately predict the interaction parameters for the Hamiltonians simulated by quantum gas microscopes. 
Hubbard-model simulations 
rely on the precise determination of the hopping parameter $t$ and the on-site interaction energy $U$.
While the hopping $t$ can be accurately determined from a single-particle picture, the calculation of $U$ is, in general, a challenging task when 
the underlying short-ranged atom-atom interactions are strong, i.e., the magnitude of the $s$-wave scattering length $a$ becomes comparable to the length scales 
associated with the optical potentials, such as the lattice spacing $d$ and transverse harmonic oscillator length $l$. 
The most commonly used theoretical approximation for $U$ is to replace the $s$-wave pseudopotential with a Dirac-delta interaction and then 
restrict the two-atom system to the lowest energy band~\cite{jakschColdBosonicAtoms1998}.
However, this perturbative approach ignores the short-distance behavior of the interactions 
and associated scattering into higher energy bands/levels, 
and is thus limited to weak interactions $|a|\ll d,l$~\cite{buchlerMicroscopicDerivationHubbard2010}. 
Alternatively, one can apply the exact solution for two particles in a harmonic oscillator~\cite{buschTwoColdAtoms1998} at each lattice site, thus properly accounting for virtual excitations to higher energy levels, 
but this is limited to very deep lattices~\cite{dickerscheidFeshbachResonancesOptical2005,dienerFermionsOpticalLattices2006,woutersTwobodyProblemPeriodic2006}. 

One ingenious approach that addresses these limitations is to extract $U$ by 
equating the exact two-body scattering amplitude to that of the Hubbard model in the low-energy regime, and this has already been successfully implemented in the case of a three-dimensional (3D) cubic optical lattice~\cite{buchlerMicroscopicDerivationHubbard2010}. 
However, a similar solution for quasi-one-dimensional (quasi-1D) and quasi-two-dimensional (quasi-2D) lattices is currently lacking, despite the proliferation of low-dimensional optical-lattice experiments and quantum gas microscopes.  
While there has recently been a series of works that investigated this problem in reduced dimensions~\cite{zhangKondoEffectAlkalineearthmetal2016,chengEnhancingKondoCoupling2017,zhangControlSpinexchangeInteraction2018,zhangTightbindingKondoModel2020}, these were limited to 1D and quasi-1D systems and only approximately modelled the motion in either the optical lattice or the confining harmonic potential. Thus, a complete calculation for $U$ remains an outstanding problem.
\begin{figure}
    \centering
     \includegraphics[width=\linewidth]{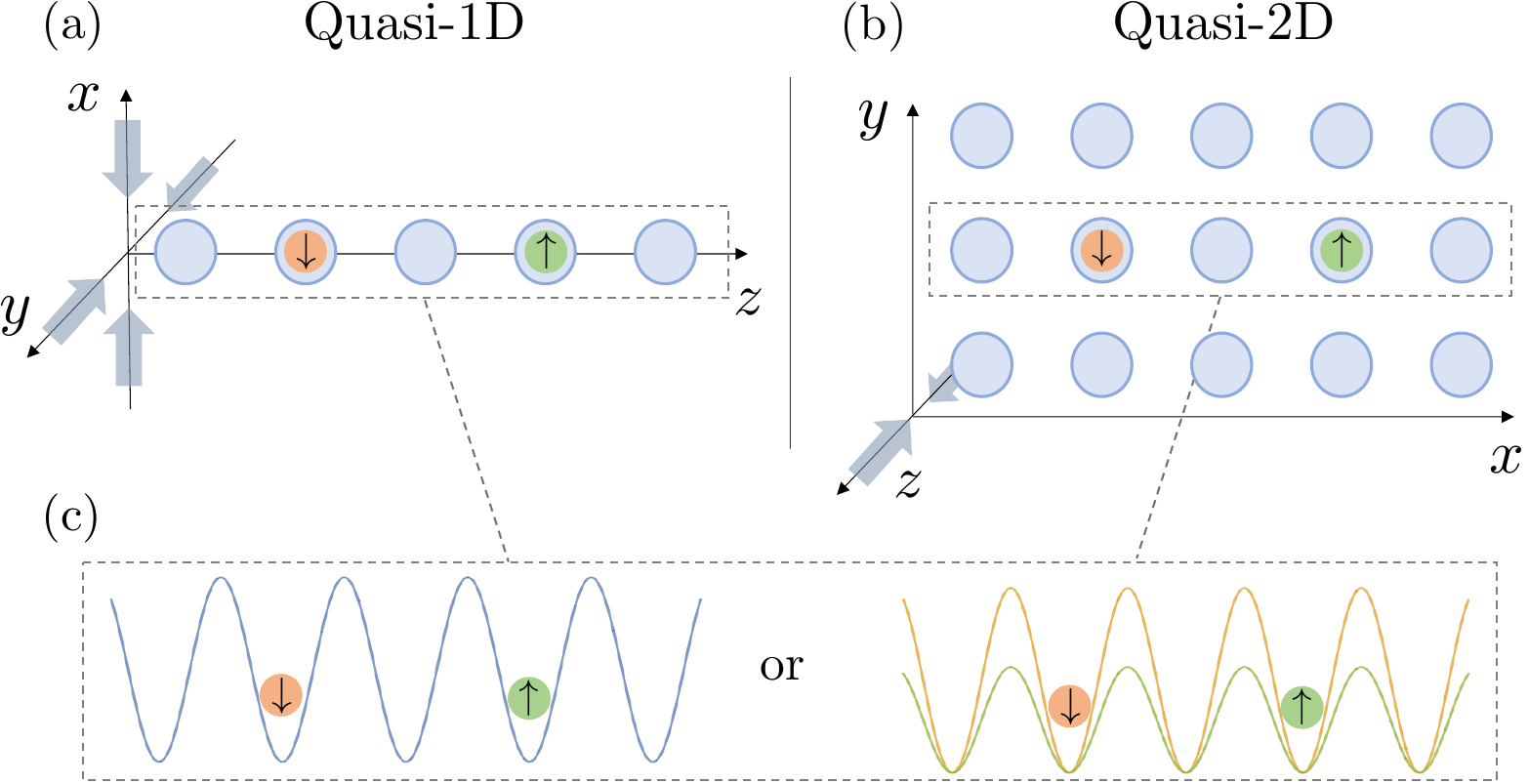}
    \caption{Illustration of the low-dimensional geometries under consideration for two interacting atoms.  
    (a) In quasi-1D, we have an optical lattice along the $z$ direction, with a transverse cylindrically symmetric harmonic confinement (i.e., along the $x$ and $y$ directions). (b) In quasi-2D, we instead consider a square lattice in the $x$-$y$ plane, with transverse harmonic confinement along the $z$ direction. The lattice strength can differ along the $x$ and $y$ directions. (c) Along a particular axis, the two atoms can either experience an identical potential (left) or a state-dependent lattice (right, where the orange spin-$\downarrow$ atom experiences the orange potential). We will use the ``dimensions of confinement'' to refer to $x,\, y$ in quasi-1D and to $z$ in quasi-2D. Meanwhile, the ``dimensions of the lattice'' refer to $z$ in quasi-1D and to $x,\, y$ in quasi-2D.}
    \label{fig:schematic}
\end{figure}

In this work, we provide an exact solution to the problem of two atoms ($\uparrow, \downarrow$) interacting in a quasi-1D or quasi-2D optical lattice, taking into account the transverse harmonic confining potential, which is relevant for experiments. 
For the sake of generality, we consider the lattice to depend on the atoms' internal state, and---in the case of a 2D lattice---to have directional dependence, as schematically shown in Fig.~\ref{fig:schematic}. 
While we focus on square lattices in 2D, our results can be easily generalized to other 2D lattice geometries such as the triangular lattice and the multi-band Lieb lattice.

Inspired by the approach of Ref.~\cite{buchlerMicroscopicDerivationHubbard2010}, we determine the Hubbard on-site interaction $U$ at arbitrary atom-atom interactions by enforcing that the Hubbard model reproduces the exact low-energy scattering amplitude for two particles in the lowest Bloch band. 
We find that $U$ exhibits broad resonances for scattering lengths comparable to the lattice spacing, which is reminiscent of confinement-induced resonances in a uniform quasi-1D geometry~\cite{olshaniiAtomicScatteringPresence1998}.
We also determine the exact two-body bound-state energies and we find that they remain finite when $U$ diverges, a feature which is absent in the single-band Hubbard model. 
Thus, like the confinement-induced resonance in quasi-1D, the divergence in $U$ is due to the presence of higher energy bands~\cite{Bergeman2003}. While our effective Hubbard $U$ cannot capture the bound states for arbitrary scattering length $a$, we expect it to provide an accurate description of the low-energy scattering properties, such as might be found in a repulsive many-body state.
For instance, confinement-induced resonances have already been successfully used to produce the correlated Tonks-Girardeau gas in 1D~\cite{paredesTonksGirardeauGas2004}.  
Finally, we compare our calculation of $U$ with the experimental determination of $U$ via lattice modulation spectroscopy in a quasi-2D lattice~\cite{greifSiteresolvedImagingFermionic2016} and find excellent agreement.

This paper is organized as follows. In Sec.~\ref{Sect:Model}, we introduce the microscopic model of two atoms interacting in an optical lattice in the presence of a transverse harmonic confinement. Section~\ref{Sect:Renormalization} is devoted to the formal solution of the two-body problem, where we provide significant details on the renormalization of the model. Furthermore, we calculate the exact two-body $T$ matrix, which we relate to the scattering amplitude of two atoms in the lowest Bloch band. In Sec.~\ref{Sect:HubbardParameters}, we show how the on-site interaction energy $U$ can be calculated by equating the Hubbard and the exact scattering amplitudes, and we compare this method with perturbative approaches. We present our numerical results for $U$, as well as for the two-body bound-state energies. We investigate the ability of the Hubbard model to reproduce these bound states, and finally compare with experiment. We conclude in Sec.~\ref{Sect:conclusion}.
Technical details on the model, the numerical implementation of the $T$ matrix, the on-shell $T$ matrix calculation, and the determination of effective masses can be found in Appendix~\ref{Appendix:TwoChannelHam}, \ref{Appendix:NumImplement}, \ref{Appendix:OnShellTMat} and \ref{Appendix:DetEffectiveMass}, respectively. Our source code is available on GitHub~\cite{SourceCode}.

\section{Model} \label{Sect:Model}
We consider the problem of two interacting atoms confined to an optical lattice in low dimensions, as illustrated in Fig.~\ref{fig:schematic}. Here we focus on two scenarios: a regular lattice in a quasi-1D geometry and a square lattice in a quasi-2D geometry.
We take the two atoms to have the same mass $m$, but to be potentially distinguishable via their internal hyperfine state, which we denote $\sigma =\uparrow, \,\downarrow$. Note that our results for the Hubbard interaction do not depend on whether the two distinguishable atoms are fermionic or bosonic, and they also apply to two identical bosons as long as the external potentials are identical. 

We use the most common optical lattice potentials:
\begin{subequations}
\label{eq:lattice}
\begin{align}
    V^{\rm 1D}_\sigma(z) &= v^z_{\sigma} \sin^2 (\pi z/d),\\
    V^{\rm 2D}_\sigma(x,y) &= v^x_{\sigma} \sin^2 (\pi x/d) + v^y_{\sigma} \sin^2 (\pi y/d),
\end{align}
\end{subequations}
where $v^{i}_{\sigma}$ is the lattice depth along the $i$ direction and $d$ is the lattice spacing. The natural energy scale for the lattice is the recoil energy $V_r \equiv \frac{\pi^2}{2 md^2} $ (we work in units in which $\hbar$ and the system volume are set to unity). For the sake of generality, we allow the lattice depth to depend on both the atom's hyperfine state (``spin'') and the direction.

The atoms are also confined in the direction transverse to the optical lattice by a harmonic oscillator (HO) potential:
\begin{subequations}
\label{eq:HO}
\begin{align}\label{Eq:HOQuasi1D}
    V_{\rm HO}^{\rm 1D}(x,y) &= \frac{1}{2} m \omega^2 (x^2 + 
    y^2),\\ \label{Eq:HOQuasi2D}
    V_{\rm HO}^{\rm 2D} (z) &= \frac{1}{2} m \omega^2 z^2,
\end{align}
\end{subequations}
where $\omega$ is the oscillator frequency, which is taken to be independent of the atom spin. For the quasi-1D geometry, the confining potential in Eq.~\eqref{Eq:HOQuasi1D} corresponds to a 2D harmonic oscillator; thus we index the HO eigenstates by quantum number $\sv = \{s ,j\}$ with corresponding HO eigenenergy $s \omega$, where $s$ is the radial quantum number, while $j$ is the 2D angular momentum quantum number. Likewise, in the quasi-2D geometry in Eq.~\eqref{Eq:HOQuasi2D}, we have a 1D harmonic oscillator with  $\sv = \{s  \}$, where the associated HO eigenenergy is also $s \omega$ (but where the energy levels are non-degenerate). Here, and throughout, we measure all energies relative to the zero-point HO energy.

\subsection{Hamiltonian}

We now introduce the  Hamiltonian that describes the lattice and transverse confinement in Eqs.~\eqref{eq:lattice} and \eqref{eq:HO}, as well as interactions between the two atoms. The Hamiltonian consists of four terms
\begin{align} \label{Eq:CompleteHamiltonian}
    \Hh = \Hh_\uparrow + \Hh_\downarrow + \Hh_{\rm c} + \Hh_{\rm{co}}, 
\end{align}
where the first two terms capture the atoms in the absence of interactions, while the last two terms describe their interactions. Since the basic formalism is the same for the quasi-1D and quasi-2D geometries, we will present them both together using a unified notation in which the dimensions of vectors, sums and indices are implicit. 

The single-particle terms in the Hamiltonian are
\begin{align} \label{Eq:Hsigma}
    \Hh_\sigma = \sum_{\Kv\sv} \varepsilon_{\Kv \sv} \chd_{\Kv \sv \sigma} \ch_{\Kv \sv \sigma} + \sum_{\Kv \Qv \sv} \tilde{V}_\sigma(\Qv) \chd_{\Kv+\Qv,\sv \sigma} \ch_{\Kv  \sv \sigma},
\end{align}
where $\tilde{V}_\sigma(\Qv)$ is the Fourier transform of the optical lattice potential and $\chd_{\Kv \sv \sigma}$ creates a spin-$\sigma$ atom with momentum $\Kv$, HO quantum number $\sv$, mass $m$ and energy $\varepsilon_{\Kv s} = \epsilon_{\Kv}+s\omega$, with $\epsilon_{\Kv} = |\Kv|^2/2m\equiv K^2/2m$. Here, the momentum is along the direction(s) of the 1D or 2D lattices, while the HO quantum number denotes the single-particle eigenstates transverse to the optical lattice (see Fig.~\ref{fig:schematic}). 

According to Bloch's theorem, the single-particle eigenstates of $\hat H_\sigma$ satisfy
\begin{align} \label{Eq:BlochStates}
    \Hh_\sigma \ket{\kv, \nuv, \sv, \sigma} = (E^\sigma_{\kv \nuv} + 
    s \omega)\ket{\kv, \nuv, \sv, \sigma},
\end{align}
where $\kv$ is now the quasimomentum with components along each lattice direction restricted to lie within the first Brillouin zone, i.e., $k_i \in (-\pi/d, \pi/d]$, while $\nuv$ is the band index (with components $\nu_i =0, 1,2,\dots$~\footnote{In contrast to standard treatments of lattices, we employ a vector for the band index in the quasi-2D geometry, which is convenient owing to the separability of the square lattice along the $x$ and $y$ directions.}). Here, and throughout, we use capital and lower case letters for regular momenta and quasi-momenta, respectively. In Eq.~\eqref{Eq:BlochStates} we have defined
\begin{align}\label{eq:BlochStates2}
    \ket{\kv, \nuv, \sv, \sigma} \equiv \sum_{\nv} \varphi^{(\kv,\nuv,\sigma)}_{\nv} \chd_{\kv + 2\pi \nv/d ,\sv \sigma} \ket{0},
\end{align}
where $\nv$ has components $n_i =0, \pm 1, \pm2, \dots$ along each lattice direction and $\ket{0}$ is the vacuum. Since the different dimensions are separable at the single-particle level, the amplitudes can be written as $\varphi^{(\kv,\nuv,\sigma)}_{\nv} \equiv \prod_{i} \varphi^{(k_i,\nu_i,\sigma)}_{n_i}$, where the different components satisfy
\begin{align} \label{Eq:ExpansionEqAtoms}
    E^{\sigma}_{k_i \nu_i} \varphi_{n_i}^{(k_i,\nu_i,\sigma)} &= \left( \epsilon_{k_i,n_i} + \frac{v^i_\sigma}{2}\right) \varphi_{n_i}^{(k_i,\nu_i,\sigma)} \nonumber\\&{} \quad - \frac{v^i_\sigma}{4} \left(\varphi_{n_i-1}^{(k_i,\nu_i,\sigma)} + \varphi_{n_i+1}^{(k_i,\nu_i,\sigma)} \right) .
\end{align}
Here, $E^{\sigma}_{k\nu} $ are the energy eigenvalues along the corresponding direction such that $E^{\sigma}_{\kv \nuv} = \sum_{i }E^{\sigma}_{k_i\nu_i}$, and $\epsilon_{k,n} = (k+ 2\pi n/d)^2/2m$. Note that Eq.~\eqref{Eq:ExpansionEqAtoms} exhibits
parity symmetry such that $E^\sigma_{k,\nu} = E^\sigma_{-k,\nu}$ and $| \varphi_n^{(k,\nu,\sigma)}| =  | \varphi_{-n}^{(-k,\nu,\sigma)}|$. The latter relation implies that we can define the phase such that $\varphi_{n}^{(k,\nu,\sigma)}=(-1)^\nu \varphi_{-n}^{(-k,\nu,\sigma)}$ without loss of generality, which is useful in the coming analysis of symmetries.

The last two terms in the Hamiltonian \eqref{Eq:CompleteHamiltonian} capture how the atoms interact via a Feshbach resonance. Similarly to the calculation of Hubbard parameters for a 3D lattice~\cite{buchlerMicroscopicDerivationHubbard2010}, we model this using a two-channel Hamiltonian~\cite{timmermansFeshbachResonancesAtomic1999}, which is convenient since it simplifies the description of center-of-mass  and relative motion within the lattice. In this model, the atoms interact by converting into a closed-channel diatomic molecule which, in our geometry, is described by the Hamiltonian
\begin{align} 
    \Hh_{\rm c} = &\sum_{\Kv \Sv} \left( \frac{\varepsilon_{\Kv, 2 S}}{2} + \delta_D\right) \dhd_{\Kv \Sv} \dh_{\Kv \Sv} \nonumber \\
    &{}\qquad +\sum_{\Kv \Qv \Sv}
    \tilde V_\mathrm{tot}(\Qv)  \dhd_{\Kv+\Qv,\Sv} \dh_{\Kv\Sv},\label{Eq:ClosedChannelHam}
\end{align}
where $\dhd_{\Kv \Sv}$ creates a molecule with momentum $\Kv$, HO quantum number $\Sv$, mass $2m$ and energy $\varepsilon_{\Kv,2S}/2 + \delta_D$. Here, $\delta_D\equiv\delta-\omega/D$ is defined in terms of the bare closed-channel detuning in 3D, $\delta$, and we have taken into account the effective reduction of the detuning due to the zero-point energy of the relative motion, with $D=1$ and 2 in the quasi-1D and quasi-2D geometries, respectively. 

The molecule, corresponding to the center-of-mass (CM) motion of the atoms, experiences the total lattice potential $\tilde V_\mathrm{tot}(\Qv)\equiv \tilde V_\uparrow(\Qv)+\tilde V_\downarrow(\Qv)$, as well as the transverse harmonic confinement of frequency $\omega$. Similarly to the atoms, the single-molecule part of the Hamiltonian can be diagonalized to obtain 
\begin{align}\label{eq:knumstates}
    \Hh_{\rm{c}} \ket{\kv, \nuv, \Sv, M}  = (E^M_{\kv \nuv}  + S\omega+ \delta_D)\ket{\kv, \nuv, \Sv,M},
\end{align}
with quasimomentum $\kv$ and band index $\nuv$. Here, we have introduced
\begin{align}
    \ket{\kv,\nuv,\Sv,M} = \sum_{\nv} \eta_{\nv}^{(\kv,\nuv)} \dhd_{\kv + 2\pi \nv/d,\Sv} 
    \ket{0},
\end{align}
where the amplitudes $\eta_{\nv}^{(\kv,\nuv)} \equiv \prod_i \eta_{n_i}^{(k_i,\nu_i)}$ and the energy eigenvalues $E^M_{\kv \nuv} \equiv \sum_i E^M_{k_i \nu_i}$ satisfy
\begin{align}
    E^M_{k_i \nu_i} \eta_{n_i}^{(k_i,\nu_i)} &= \left( \frac{\epsilon_{k_i,n_i}}{2} + \frac{v^i_\uparrow + v^i_\downarrow}{2} \right) \eta_{n_i}^{(k_i,\nu_i)} \nonumber
    \\ \label{eq:ExpansionEqClosed}
    &{}\quad - \frac{v^i_\uparrow + v^i_\downarrow}{4} \left( \eta_{n_i-1}^{(k_i,\nu_i)} +  \eta_{n_i+1}^{(k_i,\nu_i)}  \right).
\end{align}
Once again, $E^M_{k, \nu} = E^M_{-k, \nu}$ and $\eta_n^{(k,\nu)}=(-1)^\nu\eta_{-n}^{(-k,\nu)}$ due to the symmetry of the lattice.

We will assume that the process by which the atoms form the closed-channel molecule is of effectively zero range, in which case the corresponding term in the Hamiltonian takes the following form in the presence of a transverse harmonic confinement
\begin{align} \label{Eq:InteractionHamiltonian}
    \Hh_{\rm{co}} = g \sum_{\substack{\Kv \Qv \\ \sv_1 \sv_2 \Sv}} \xi^{\Sv \Kv}_{\sv_1 \sv_2} \dhd_{\Qv \Sv} \ch_{\Qv/2+\Kv,\sv_2\downarrow} \ch_{\Qv/2-\Kv,\sv_1\uparrow} + \rm{h.c.},
\end{align}
where $g$ is the 3D coupling strength between open and closed channels --- see Appendix~\ref{Appendix:TwoChannelHam} for details on how to derive Eq.~\eqref{Eq:InteractionHamiltonian} from the usual two-channel interaction in 3D.
The form factor $\xi^{\Sv \Kv}_{\sv_1 \sv_2}$ involves the change of basis from the individual HO quantum numbers $\{\sv_1, \sv_2\}$ to those for the CM and relative motion $\{\Sv,\sv\}$, giving
\begin{align}\label{eq:CG}
\xi^{\Sv \Kv}_{\sv_1 \sv_2}= 
\sum_{\sv} \chi(s,K) \, \phi_{\sv}
\braket{\Sv,\sv}{\sv_1,\sv_2} ,
\end{align}
where $
\braket{\Sv,\sv}{\sv_1,\sv_2}$ is the Clebsch-Gordan coefficient for the basis transformation~\footnote{As will become apparent, we never actually need the precise form of these. However, they have previously been obtained by explicit evaluation in Ref.~\cite{Smirnov1962}, and in terms of Wigner's $d$ matrix using the mapping between a 2D harmonic oscillator and angular momentum~\cite{Levinsen2014PRX}.} 
and $\chi(s,K)$ is a function that regularizes the divergent ultraviolet physics associated with the relative motion for a zero-range interaction (discussed in detail in Sec.~\ref{Sect:Renormalization} below). Furthermore, the coefficient $\phi_{\sv} \equiv \phi_{\sv} (\xv=\zerov)$, where $\phi_{\sv}(\xv)$ is the real-space HO eigenfunction in the relative frame (i.e., for a particle of mass $m/2$ in a harmonic oscillator of frequency $\omega$). 
To be specific, in quasi-1D where $\sv = \{s,j \}$
\begin{align} \label{eq:phiq1D}
    |\phi_{\sv}|^2=\left\{\begin{array}{cl} \frac1{2\pi l^2} & s \text{ even and } j=0 \\[4pt] 0
    & \mathrm{otherwise} \end{array}\right.,
\end{align}
whereas in quasi-2D
\begin{align} \label{eq:phiq2D}
    |\phi_{s}|^2=\left\{\begin{array}{cl} \frac{1}{\sqrt{2\pi l^2}}\frac1{2^{s}}\binom{s}{s/2} & s \text{ even} \\[4pt] 0
    & \mathrm{otherwise} \end{array}\right..
\end{align}
In both expressions, $l\equiv 1/\sqrt{m \omega}$ is the HO length.

\section{Two-body problem}
\label{Sect:Renormalization}

In 3D uniform space, the low-energy scattering amplitude for two distinguishable atoms (or two identical bosonic atoms) with short-range interactions is~\cite{landau2013quantum}
\begin{align} \label{Eq:3DLowEnergyScatteringAmplitude}
    f(K_{3\rm{D}}) = - \frac{1}{a^{-1} - \new{r_e} K_{3\rm{D}}^2/2 +iK_{3\rm{D}}},
\end{align}
where $\kvt$ is the 3D  relative momentum, $a$ is the 3D $s$-wave scattering length and 
\new{$r_e$ is the effective range. We consider the scenario near a Feshbach resonance, where $r_e$ is negligible in the case of a broad resonance, or negative in the case of a narrow one.  
Thus, it is convenient to instead use the (positive) range parameter $R^* \equiv -r_e/2$.} 
The physical parameters ($a$, $R^*$) can in turn be related to the bare two-channel parameters ($g$, $\delta$) via the process of renormalization. Specifically, we require that the scattering amplitude of our model in the absence of any confining potentials reproduces the low-energy behavior of Eq.~\eqref{Eq:3DLowEnergyScatteringAmplitude}, which yields the renormalization equations~\cite{gurarieResonantlyPairedFermionic2007}
\begin{subequations}
\begin{align}
    \frac{m}{4 \pi a} &= - \frac{\delta}{g^2} + \sum_{\kvt}\chi^2_{\rm 3D}(K_{\rm 3D})\frac{1}{2\epsilon_{\kvt}}, \label{Eq:Renorm3D}{}\\
    R^*&= \frac{4\pi}{m^2g^2},
\label{Eq:Renorm3DRstar}
\end{align}
\end{subequations}
where we have introduced a cutoff function $\chi_{\rm 3D}$ which regularizes the ultraviolet (UV) divergence. Eventually, we will take the UV cutoff to infinity such that our results are independent of the UV physics. Note that the two-channel model is equivalent to a standard single-channel model of \new{a broad Feshbach resonance} in the limit of $\delta, g \to \infty$, provided one keeps $u_{\rm 3D} = - g^2/\delta$ constant, where $u_{\rm 3D}$ defines the (bare) coupling constant of the contact interactions between the two atoms
\footnote{To be precise, in the single-channel case the two atoms interact via the Hamiltonian 
\begin{align*}
    \hat H=u_\mathrm{3D}\sum_{\kvt \kvt' \qvt}\chd_{\kvt,\downarrow} \chd_{\kvt',\uparrow}\ch_{\kvt'+\qvt,\uparrow}\ch_{\kvt-\qvt,\downarrow}.
\end{align*}
The limit of the single-channel Hamiltonian can be formally arrived at from the two-channel Hamiltonian $\Hh_\mathrm{c}+\Hh_\mathrm{co}$ by taking the limit $\delta, g \to \infty$, where one identifies the ratio $u_{\rm 3D} = - g^2/\delta$.}. This corresponds physically to taking \new{$R^*\to 0$}, while keeping the scattering length finite.

The challenge now is to obtain properly renormalized expressions for the low-energy scattering properties in the presence of both harmonic confinement and the optical lattice. 
While we could, in principle, renormalize our calculation by employing Eq.~\eqref{Eq:Renorm3D} directly, it is more convenient to first re-express it in a form that accounts for the strong harmonic confinement, thus describing the effective dimensionality of the atom-atom scattering. 
We can then exploit the known exact solutions for the scattering parameters in quasi-1D and quasi-2D geometries in the absence of a lattice --- see Refs.~\cite{olshaniiAtomicScatteringPresence1998} and \cite{petrovInteratomicCollisionsTightly2001}, respectively. 
Therefore, we will start by deriving the low-dimensional versions of Eq.~\eqref{Eq:Renorm3D} in Sec.~\ref{sec:lowD}  before turning to the effects of the lattice in Sec.~\ref{Sect:OpenChannelTMatrix}.

In the following sections, we use the open-channel $T$ matrix 
to determine the low-energy scattering properties of the two-atom problem. 
The corresponding $T$ operator is obtained from  the closed-channel Green's operator via
\begin{align} \label{Eq:TMatrixFromD}
    \Th(E+i0) = g^2 \Hh_{\rm{int}} \Dh(E+i0) \Hh_{\rm{int}},
\end{align}
where $g \Hh_{\rm{int}} \equiv \Hh_{\rm{co}}$ and $i0$ is an infinitesimal positive shift into the complex energy plane. Here, the closed-channel Green's operator is given by
\begin{subequations}\label{eq:closedchG}
\begin{align}
    \Dh(E) &= \Dfreeh\!(E)+ g^2 \Dfreeh\!(E) \Pih (E) 
    \Dfreeh\!(E) + \dots \label{Eq:SuppressedEnergy} \\
    &=\frac{1}{[\Dfreeh\! (E)]^{-1} - g^2 \Pih (E)},
\end{align}
\end{subequations}
where
\begin{align}
    \Pih(E) \equiv \Hh_{\rm{int}} \Gfreeh\! (E) \Hh_{\rm{int}},
    \label{eq:Pih}
\end{align}
is the polarization bubble while
\begin{subequations}
\label{eq:freegreen}
    \begin{align}
        \Gfreeh\!(E) &= (E - \Hh_{\uparrow} - \Hh_{\downarrow})^{-1},\\
        \Dfreeh\!(E) &= (E - \Hh_{\rm{c}})^{-1},
    \end{align}
\end{subequations}
are the free open- and closed-channel Green's operators, respectively. 

\subsection{Renormalization in low dimensions} \label{sec:lowD}
Before tackling the full two-particle problem in a lattice, we first determine the effect of the low-dimensional geometry on the scattering properties. 
To this end, we set $v^i_\sigma=0$ and we consider the scattering of two particles along the dimension(s) perpendicular to the harmonic confinement. Since the CM and relative motion separate along all dimensions in the absence of a lattice, we may take the CM momentum $\Qv$ and the CM HO quantum number $\Sv$ to be zero, without loss of generality. We then consider the state of two particles at relative momentum $\Kv$ and relative HO quantum number $\sv$,
\begin{align}
    \ket{\Kv,\sv} = \sum_{\sv_1,\sv_2} 
    \braket{\sv_1,\sv_2}{\Sv=0,\sv}
    \chd_{-\Kv, \sv_1, \uparrow}  \chd_{\Kv, \sv_2, \downarrow} \ket{0}.
\end{align}
In this basis, the relevant matrix element of the interaction is simply (for details, see Appendix~\ref{Appendix:TwoChannelHam})
\begin{align} \label{eq:intmatrixelt}
    \bra{0} \dh_{\zerov\zerov} \Hh_{\rm int} \ket{\Kv,\sv} = \phi_{\sv} \chi(s, K).
\end{align}
Importantly, from Eqs.~\eqref{eq:phiq1D} and \eqref{eq:phiq2D}, we see that the matrix element is only non-zero when $s$ is even in  the quasi-2D geometry, and for $s$ even \textit{and} $j=0$ in the quasi-1D geometry.

We are now in a position to determine the appropriate regularization function $\chi$. While there are a number of possible choices, we must remove the ultraviolet behavior in at least two of the three dimensions, since zero-range interactions are well defined in one dimension. In both quasi-1D and quasi-2D, we employ a simple function that is unity at low energy while cutting off the high-energy physics related to the motion in the $x$-$y$ plane, corresponding to
%
\begin{align}
\label{Eq:OurRegularization}
    \chi(s,K) = \bigg\{\begin{array}{ccc}
    \chi_1(s) &= \Theta(\Lambda' - s), & \text{quasi-1D}\\
    \chi_2(K) &= \Theta(\Lambda - K),& \text{quasi-2D} 
    \end{array},
\end{align}
%
where $\Theta$ is the Heaviside function. We have introduced $\Lambda' = \Lambda^2/2m\omega$ to ensure the same energy cutoffs in both geometries. Our choice of cutoff is convenient since it only involves either the dimensions of confinement (quasi-1D) or the dimensions of the lattice (quasi-2D), but it does not involve both the confinement and the lattice at the same time.

The low-energy scattering is captured by the quasi-1D or quasi-2D $T$ matrix $\mathcal{T}_\mathcal{Q}(E) \equiv \bra{\Kv,\sv=0} \Th(E) \ket{\Kv, \sv=0}$ (note that this does not depend on $\Kv$ due to the use of a zero-range interaction). By using the definition~\eqref{Eq:TMatrixFromD} together with the matrix elements in Eq.~\eqref{eq:intmatrixelt} we obtain 
\begin{align}\label{eq:TQ}
    \mathcal{T}_\mathcal{Q}^{-1}(E)=  \frac{1}{|\phi_0|^2}\left( \frac{E-\delta_D}{g^2} - \Pi_\mathcal{Q} (E)\right).
\end{align}
For low energies where $E < 2\omega$,
the matrix element of the polarization bubble in Eq.~\eqref{eq:Pih} is given by
\begin{align}
\nonumber
    \Pi_\mathcal{Q}(E) &= \sum_{\Kv} \sum_{s=0} \chi^2(2s,K)|\phi_{2s}|^2  \frac{1}{E- 2\varepsilon_{\Kv,s} + i0}\\
    &\hspace{-10mm} = |\phi_0|^2 \sum_{\Kv} \frac{\chi^2(0,K)}{E- 2\epsilon_{\Kv}+i0} 
    - \sum_{s=1}|\phi_{2s}|^2\sum_{\Kv}   \frac{\chi^2(2s,K)}{-E+ 2\varepsilon_{\Kv,s}},
\label{eq:polbubble}
\end{align}
%
where we have used the fact that $\phi_s=0$ for odd $s$. In the case of a quasi-1D geometry, we have taken the quantum number $j=0$ and defined $\phi_s \equiv \phi_{\{s,0\}}$. 


In the last step of Eq.~\eqref{eq:polbubble}, we have separated out the $s=0$ component of $|\phi_0|^{-2}\Pi_\mathcal{Q}$, since this corresponds to the polarization bubble of a pure 1D/2D geometry. In particular, ignoring the dimensions of confinement, the low-energy, pure 1D/2D scattering is captured by the $T$ matrix element $\mathcal{T}_\mathcal{P}(E) \equiv \bra{\Kv} \Th(E) \ket{\Kv}$, which is given by
\begin{align} \label{Eq:RenormTMatrixPure}
    \mathcal{T}_\mathcal{P}^{-1}(E) = \frac{1}{u} - \sum_{\Kv}  \frac{\chi^2(0,K)}{E - 2\epsilon_{\Kv} + i0},
\end{align}
where $u_{\rm 1D}$ and $u_{\rm 2D}$ are the low-dimensional coupling constants of a contact interaction. We have
\begin{subequations}
\label{eq:u1dandu2d}
    \begin{align}
    \frac1{u_{\rm 1D}} & = -\sum_{\Kv}\frac{1}{\frac{1}{ma_{\rm 1D}^2}+2 \epsilon_{\Kv}}=- \frac{ma_{\rm 1D}}{2},\\
    \frac1{u_{\rm 2D}} & = -\sum_{\Kv}
    \frac{\chi^2_2(K)}{\frac{1}{ma_{\rm 2D}^2}+2 \epsilon_{\Kv}},
\end{align}
\end{subequations}
with $a_{\rm 1D}$ and $a_{\rm 2D}$ the effective 1D and 2D scattering lengths, respectively. Note that while $u_{\rm 1D}$ is finite, $u_{\rm 2D}$ requires renormalization.

Equating the ``quasi'' $(\mathcal{T}_\mathcal{Q})$ and ``pure'' ($\mathcal{T}_\mathcal{P}$) $T$ matrices in the limit $E\to 0$ yields the renormalization equations in quasi-1D and quasi-2D respectively,
\begin{subequations}\label{eq:OurRenormEq}
\begin{align} 
     \frac{\delta-\omega}{g^2}  &= \frac{m a_{\rm 1D}}{4 \pi l^2}  + \frac{1}{2 \pi l^2} \sum_{K} \sum_{s=1}
     \frac{\chi^2_1(2s)}{2 \varepsilon_{K,s}}, \label{Eq:RenormOlshanii}\\
    \frac{\delta-\omega/2}{g^2} &= \frac{1}{\sqrt{2 \pi l^2}}\sum_{\Kv}\!
    \frac{\chi^2_2(K)}{\frac{1}{ma_{\rm 2D}^2}+2 \epsilon_{\Kv}} \!+\!\sum_{\Kv}\!
    \sum_{s=1} |\phi_{2s}|^2 \frac{\chi^2_2(K)}{ 2 \varepsilon_{\Kv,s}}.
    \label{Eq:RenormPetrov}
\end{align}
\end{subequations}
The single-channel limit ($\omega/g^2 \to 0$) 
of Eq.~\eqref{Eq:RenormOlshanii} was originally obtained in Ref.~\cite{olshaniiAtomicScatteringPresence1998}, and likewise the single-channel limit of Eq.~\eqref{Eq:RenormPetrov} was obtained in Ref.~\cite{petrovInteratomicCollisionsTightly2001} albeit in a different formulation.

Finally, the effective scattering lengths can be related to the 3D scattering parameters by comparing the renormalization equations in Eq.~\eqref{eq:OurRenormEq} to the 3D equation in Eq.~\eqref{Eq:Renorm3D} with $\chi_{\rm 3D}(\kvt) \to \chi(s, \Kv)$. 
Taking the cutoff to infinity and using the known procedure for the case of single-channel interactions~\cite{olshaniiAtomicScatteringPresence1998,petrovInteratomicCollisionsTightly2001}, we obtain
\begin{subequations}\label{eq:a1da2d}
\begin{align}
    a_{\rm 1D} &=  
    - l \left( \frac{l}{a} +\frac{R^*}{l} + \frac{\zeta(1/2)}{\sqrt{2}} \right),\\
    a_{\rm 2D} &= l \sqrt{\frac{\pi}{B}} \exp\left[- \sqrt{\frac{\pi}2}\left(\frac l a+\frac{R^*}{2l}\right)\right], \quad B \approx 0.905,\label{eq:a2dkirk}
\end{align}
\end{subequations}
where $\zeta$ is the Riemann zeta function. The two-channel correction in Eq.~\eqref{eq:a2dkirk} was derived in Ref.~\cite{Kirk2017}.

Inserting Eqs.~\eqref{eq:polbubble} and \eqref{eq:OurRenormEq} in Eq.~\eqref{eq:TQ} allows us to derive the fully renormalized $T$ matrix in the confined geometry in terms of the 1D and 2D scattering lengths. These can in turn be related to the 3D scattering parameters $a$ and $R^*$ using Eq.~\eqref{eq:a1da2d}. We will now use this procedure to replace bare parameters with fully renormalized quantities for the two-particle problem in a lattice.

\subsection{Scattering in the presence of a lattice}
\label{Sect:OpenChannelTMatrix}
We now construct the exact solution of the two-body problem in the presence of a lattice. 
As seen in Eq.~\eqref{Eq:TMatrixFromD}, the $T$ matrix can be calculated from the closed-channel Green's function $\Dh$. Since the interactions preserve CM quasimomentum and 
HO quantum number, we calculate the matrix elements of the inverse of $\Dh$ at fixed quasimomentum $\qv$ and $\Sv=0$. Letting $\ket{\qv, \nuv} \equiv \ket{\qv,\nuv,\Sv=0,M}$ and using Eqs.~\eqref{eq:knumstates} and (\ref{eq:closedchG}-\ref{eq:freegreen}), we find 
\begin{subequations}
\label{eq:DGreen}
\begin{align} 
    D^{-1}_{\nuv \nuv'}(\qv,E) & \equiv \bra{\qv, \nuv} \Dh^{-1}(E) \ket{\qv, \nuv'} \\ &= \left(E - E^M_{\qv\nuv} - \delta_D \right) \delta_{\nuv\nuv'} - g^2 \Pi^{\qv}_{\nuv \nuv'},
\end{align}
\end{subequations}
where $\Pi^{\qv}_{\nuv \nuv'} \equiv \bra{\qv, \nuv} \Pih(E) \ket{\qv, \nuv'}$ and $\delta_{\nuv\nuv'}$ is the Kronecker delta. Before inserting a complete set of atom-atom states to determine the matrix element of $\Pih$, we 
note that 
the accessible two-atom states are limited to the subspace of states of the form
\begin{align}
    \ket{\kv, \qv; \nuv_1, \nuv_2; \sv}  &= \sum_{\sv_1, \sv_2}
    \braket{\sv_1,\sv_2}{\Sv=0,\sv}
    \ket{\qv/2-\kv,\nuv_1,\sv_1,\uparrow} \nonumber \\ &{}\hspace{4em} \otimes \ket{\qv/2+\kv,\nuv_2,\sv_2,\downarrow},
    \label{eq:2atomstates}
\end{align}
which is written in terms of the single-particle states introduced in Eq.~\eqref{eq:BlochStates2}.

By separating $\Hh_{\rm int}$ into its components along the lattice ($L$) and harmonic confinement $(C)$ dimensions --- i.e., $\Hh_{\rm int} = \Hh^L_{\rm int} \otimes \Hh^C_{\rm int}$ --- the relevant matrix elements of the interaction are
\begin{subequations} \label{Eq:MatrixElIncludingS}
\begin{align}
    \bra{\qv, \nuv} \Hh_{\rm int} &\ket{\kv, \qv; \nuv_1, \nuv_2; \sv}\nonumber\\
    &= \chi_1(s)\phi_{s} \bra{\qv, \nuv} \Hh^L_{\rm int} \ket{\kv, \qv; \nuv_1, \nuv_2}\\
    &\equiv \chi_1(s) \phi_{s} \mathcal{H}^{\qv,\nuv}_{\kv \nuv_1 \nuv_2}.
\end{align}
\end{subequations}
Here we have used the separability of the cutoff function in Eq.~\eqref{Eq:OurRegularization}, along with the matrix element along the dimensions of confinement, Eq.~\eqref{eq:intmatrixelt}. Furthermore, the cutoff function $\chi_1$ is only relevant in quasi-1D and can be safely set to 1 in quasi-2D.

The (real) matrix elements of the interaction in the dimensions of the lattice are thus
\begin{align} \label{Eq:HMatElementsExact}
    \mathcal{H}^{\qv,\nuv}_{\kv \nuv_1 \nuv_2} &= \sum_{\Nv,\nv} 
    \chi_2(
    2 \pi |\nv|/d)
    \eta^{(\qv,\nuv)}_{\Nv} \varphi^{(\qv/2-\kv,\nuv_1,\uparrow)}_{\Nv/2+\nv} \varphi^{(\qv/2+\kv,\nuv_2,\downarrow)}_{\Nv/2-\nv},
\end{align}
which depends on the possible wavefunctions of the two incident atoms and the outgoing molecule. In the sum, the index $N_i = 0, \pm1, \pm2, \dots$ while $n_i = 0, \pm 1, \pm 2,\dots$ ($n_i = \pm1/2, \pm 3/2, \dots$) for even (odd) $N_i$. As above, $\chi_2$ is only relevant in quasi-2D and it can be safely set to 1 in quasi-1D.
Note that in Eq.~\eqref{Eq:HMatElementsExact} we have approximated $\chi_2(|2 \pi \nv/d-\kv|) \simeq \chi_2(2\pi |\nv|/d)$, since the cutoff will only affect large $\nv$, at which point $|\kv|$ is negligible.

The matrix elements in Eq.~\eqref{Eq:HMatElementsExact} exhibit several symmetries when $\qv=0$, based on the previously identified symmetries of the single-particle eigenfunctions (see the discussions below Eqs.~\eqref{Eq:ExpansionEqAtoms} and \eqref{eq:ExpansionEqClosed}, respectively).
In quasi-1D, 
we have $\mathcal{H}^{0,\nu}_{k\nu_1 \nu_2} = (-1)^{\nu+\nu_1+\nu_2} \mathcal{H}^{0,\nu}_{-k, \nu_1 \nu_2}$,
and for a state-independent lattice ($v_\uparrow=v_\downarrow$), the matrix elements do not couple even and odd $\nu$ when $\nu_1=\nu_2$, i.e., $\mathcal{H}^{0,\nu}_{k\nu_1 \nu_1} = (-1)^\nu \mathcal{H}^{0,\nu}_{k\nu_1 \nu_1}$. 
The extension of these symmetries to the quasi-2D case is straightforward.

In summary, the matrix elements of $\Pih$ that appear in the closed-channel propagator \eqref{eq:DGreen} take the form
\begin{align} \label{Eq:PiElements}
    \Pi^{\qv}_{\nuv \nuv'} =&\sum_{\kv,\nuv_1,\nuv_2,s}
    |\phi_{2s}|^2 \mathcal{H}^{\qv,\nuv}_{\kv \nuv_1 \nuv_2} \frac{
    \chi_1^2(2s)}{E - E_{\kv, \qv; \nuv_1, \nuv_2;2s}}
    \mathcal{H}^{\qv,\nuv'}_{\kv \nuv_1 \nuv_2},
\end{align}
where 
$E_{\kv, \qv; \nuv_1, \nuv_2;2s} \equiv E^{\uparrow}_{\qv/2-\kv,\nuv_1} + E^{\downarrow}_{\qv/2+\kv,\nuv_2} + 2s \omega $ and we have ignored those terms that do not contribute to the interactions (i.e., odd $s$ and all $j\neq0$ angular momentum quantum numbers).
At $\qv=\zerov$, applying the symmetries of the matrix elements $\mathcal{H}$ together with the invariance of $E_{\kv, \zerov; \nuv_1, \nuv_2;2s}$ under $\kv\to-\kv$, it is seen that parity is conserved since $\Pi^{\zerov}_{\nuv \nuv'} = 0$ if any of the components of $\nuv+\nuv'$ are odd.
This implies that, at $\qv=\zerov$, the {interacting} two-body states, i.e., both bound {and scattered} states, have an associated parity.
For future discussions, it is useful to explicitly define states of fully even parity as those associated with $\nuv$ where all components are even (i.e., where $\nu$ is even in quasi-1D, while $\nu_x$ and $\nu_y$ are \textit{both} even in quasi-2D).

Finally, we can incorporate the renormalization equations, Eq.~\eqref{eq:OurRenormEq}, to find the renormalized closed-channel Green's function in Eq.~\eqref{eq:DGreen}. In quasi-1D this yields
\begin{align} \label{Eq:RenormDQuasi1D}
    D^{-1}_{\nu \nu'} (q,E) = {g^2}\left(\frac{E-E^M_{q \nu}}{{g^2}} - \frac{m a_{\rm 1D}}{4 \pi l^2} \right) \delta_{\nu \nu'} - g^2\tilde \Pi^q_{\nu \nu'},
\end{align}
where $\tilde \Pi$ is the renormalized polarization bubble:
\begin{align}
    \tilde \Pi^q_{\nu \nu'} &=  \sum_{k \nu_1\nu_2} \sum_{s=0}
    \frac{1}{2 \pi l^2}    \mathcal{H}^{q,\nu}_{k \nu_1 \nu_2} \frac{\chi^2_1(2s)}{E - E_{k, q; \nu_1, \nu_2;2s}}
    \mathcal{H}^{q,\nu'}_{k \nu_1 \nu_2}  \nonumber \\
    &{}\hspace{5em} 
    {+} \delta_{\nu \nu'} \frac{1}{2 \pi l^2} \sum_{k} \sum_{s=1}
    \frac{\chi^2_1(2s)}{2 \varepsilon_{k,s}}.
\end{align}
Likewise, in quasi-2D we find
\begin{align}
    &D^{-1}_{\nuv \nuv'} (\qv,E)=\nonumber\\
    &{}\hspace{2em}{g^2}\left(\frac{E-E^M_{\qv \nuv}}{{g^2}} - \frac{1}{\sqrt{2 \pi l^2}}
    {\frac{m}{2\pi}\ln(a_\mathrm{2D}/l)}\right) \delta_{\nuv \nuv'} -g^2\tilde \Pi^{\qv}_{\nuv \nuv'},\label{Eq:RenormDQuasi2D}
\end{align}
where
\begin{align}
    \tilde \Pi^{\qv}_{\nuv \nuv'} &= \sum_{\kv,\nuv_1,\nuv_2} \sum_{s=0}|\phi_{2s}|^2 \mathcal{H}^{\qv,\nuv}_{\kv \nuv_1 \nuv_2} \frac{1}{E - E_{\kv, \qv; \nuv_1, \nuv_2;2s}}
    \mathcal{H}^{\qv,\nuv'}_{\kv \nuv_1 \nuv_2} \nonumber\\
    &{} \quad + {\delta_{\nuv\nuv'}}
    \sum_{\kv}{\left(\frac1{\sqrt{2\pi l^2}}\frac{\chi^2_2(k)}{ 2 \epsilon_{\kv}+\omega}+
    \sum_{s=1}|\phi_{2s}|^2 \frac{\chi^2_2(k)}{ 2 \varepsilon_{\kv,s}}\right)},
\end{align}
which is fully renormalized.
Here we remind the reader that the regularization in quasi-1D is achieved using a cutoff on the relative harmonic oscillator levels $s$, while in quasi-2D it is achieved with a cutoff on the relative momenta of the particles, which also appears inside the interaction matrix elements [see Eq.~\eqref{Eq:HMatElementsExact}].

While the provided expressions for the closed-channel Green's function are exact, they remain challenging to implement numerically. We therefore provide significantly more information on their numerical implementation in Appendix~\ref{Appendix:NumImplement}.

\subsubsection{Bound states}
The exact closed-channel propagator we have derived also yields information about the two-atom bound states, since the bound-state energies correspond to the poles of the $T$ matrix, which coincide with the poles of $D$ according to Eq.~\eqref{Eq:TMatrixFromD}. 
The poles satisfy $\mathrm{det}^{-1}[D(\qv,E)]=0$, implying that $D^{-1}$ has a vanishing eigenvalue. Thus, the bound states are determined by solving the eigenvalue problems
\begin{subequations}
\label{eq:boundstates}
\begin{align}
    \frac{m a_{\rm 1D}}{4 \pi l^2} \delta_{\nu \nu'} &= {\frac{E-E^M_{q \nu}}{g^2}}\delta_{\nu \nu'}{-}\tilde\Pi^q_{\nu \nu'} 
    , \qquad\text{quasi-1D}, \\
    \frac{m\ln(a_\mathrm{2D}/l)}{(2 \pi)^{3/2} l} 
    \delta_{\nuv \nuv'}&= {\frac{E-E^M_{\qv \nuv}}{g^2}}\delta_{\nuv \nuv'}{-}\tilde\Pi^{\qv}_{\nuv \nuv'} 
    , \qquad\text{quasi-2D}.
\end{align}
\end{subequations}
For a given energy $E$ below the lowest two-atom band or within the bandgaps, we can thus obtain all the values of $a_{\rm 1D}$ or $a_{\rm 2D}$ for which a bound state exists by solving for the eigenvalues of the right hand side. In practice, in this work we will consider the single-channel limit where $g\to\infty$ such that the first term on the right hand side vanishes and $R^*=0$ according to Eq.~\eqref{Eq:Renorm3DRstar}. Then, for a given transverse confinement, the 3D scattering length is uniquely related to 
$a_{\rm 1D}$ or $a_{\rm 2D}$ via Eq.~\eqref{eq:a1da2d}.  
The procedure for finite $R^*$ \new{in the case of a narrow Feshbach resonance} is also straightforward and simply involves inserting Eq.~\eqref{eq:a1da2d} into Eq.~\eqref{eq:boundstates} a priori and solving directly for the inverse 3D scattering length at a given value of $R^*/l$.

\subsubsection{Scattering amplitude}
\label{subsub:scat}
We can also obtain the scattering amplitude from our exact calculation of the closed-channel Green's function. This involves finding the matrix elements of the $T$ operator in Eq.~\eqref{Eq:TMatrixFromD} using the matrix elements of the interaction in Eq.~\eqref{Eq:MatrixElIncludingS} and those of the closed channel propagator in either Eq.~\eqref{Eq:RenormDQuasi1D} or \eqref{Eq:RenormDQuasi2D}. 
This procedure can be carried out in complete generality. However, to make reference to the Hubbard model we will focus on low-energy scattering. One complication compared with the 3D optical lattice is that the scattering amplitude in both the quasi-1D and quasi-2D cases vanishes at zero momentum. Therefore, in the following we consider two atoms each in the lowest Bloch band, with zero CM quasimomentum ($\qv=\zerov$), harmonic oscillator index $s=0$, and a non-zero relative momentum $|\pv| \ll \pi/d$. In this case, the total energy is $E_{\pv} = E_{\pv,\zerov, \zerov,\zerov;0}$ and the on-shell $T$ matrix is
\begin{subequations}    
\begin{align}
    T(\pv)&=\bra{\pv} \hat{T}(E_{\pv}+i0)\ket{\pv}\\
    &= g^2|\phi_0|^2\mathcal{H}^{\zerov,\nuv}_{\p \zerov\zerov}D_{\nuv\nuv'}(\zerov,E_{\p}+i0)\mathcal{H}^{\zerov,\nuv'}_{\p \zerov\zerov} .
\end{align}
\end{subequations}
Here we have introduced the short-hand notation $\ket{\pv}\equiv\ket{\pv,\zerov;\zerov,\zerov;\zerov}$ in terms of the two-atom states introduced in Eq.~\eqref{eq:2atomstates}.

In the presence of the lattice, the effective masses of the $\uparrow$ and $\downarrow$ atoms are in general different from the bare mass $m$. Furthermore, we have the possibilities of different lattice strengths for the two spin components as well as a directional dependence of the quasi-2D lattice. Let us first neglect the latter possibility and consider the quasi-1D or isotropic quasi-2D scenarios. In these geometries, we can define a reduced mass $m_{\rm{eff}}$ from the effective masses such that, in the long-wavelength limit, the collision energy takes the form
\begin{align} \label{Eq:EffectiveMassLongWavelength}
    E_{\pv} \simeq E^\uparrow_{\zerov, \zerov}+E^\downarrow_{\zerov, \zerov}+ \frac{|\p|^2}{2m_{\rm{eff}}}.
\end{align}
We can then straighforwardly obtain the associated low-energy scattering amplitude, which has the same relationship to the $T$ matrix as in the absence of a lattice~\cite{Pricoupenko2011,levinsenStonglyInteractingTwo2015}:
\begin{subequations} \label{Eq:ScatFromT}
\begin{align}
        f(\p) &\simeq \frac{m_{\rm eff}}{
        i|\p|} T(\p), \\
    f(\p) &\simeq 2m_{\rm eff} T(\p),
\end{align}
\end{subequations}
in quasi-1D and 2D, respectively.

In the anisotropic case in quasi-2D, we can still carry out the above procedure. Since the dispersions in both the $x$ and $y$ directions are quadratic for both atoms, we can define effective reduced masses $m_{{\rm eff},x}$ and $m_{{\rm eff},y}$ along each direction. We then simply define a rescaled momentum and effective mass such that we have $\p'=(\sqrt{m_{\rm eff}'/m_{{\rm eff},x}}p_x,\sqrt{m_{\rm eff}'/m_{{\rm eff},y}}p_y)$ in terms of which the dispersion is isotropic. Importantly, as we discuss below, the precise choice of $m_{\rm eff}'$ drops out in our calculation of the Hubbard parameters, as it should.

\section{Hubbard parameters} \label{Sect:HubbardParameters}

\subsection{Hubbard model}

In the limit of tight confinement and a deep optical lattice ($
v_\sigma^i \gg V_r $), the exact model~\eqref{Eq:CompleteHamiltonian} maps onto the single-band Hubbard model with nearest neighbor hopping and on-site interactions,
\begin{align} \label{Eq:HubbardHam}
    \Hh = - \sum_{\langle \vb{i},\vb{j} \rangle,\sigma} t^{\vb{i}\vb{j}}_\sigma \left( \chd_{\vb{i},\sigma} \ch_{\vb{j},\sigma} + \rm{h.c.} \right)   + U \sum_{\vb{i}} \chd_{\vb{i}, \uparrow} \chd_{\vb{i}, \downarrow} \ch_{\vb{i}, \downarrow} \ch_{\vb{i}, \uparrow}\, .
\end{align}
Here, $\chd_{\vb{i},\sigma}$ creates a $\sigma$ atom at lattice site $\vb{i}$, and we have included a direction and state dependence in the hopping parameters $t_\sigma^{\vb{i}\vb{j}}$.
Note that nearest neighbor hopping requires $t_\sigma^{\vb{i}\vb{j}}=0$ if $\vb{i}$ and $\vb{j}$ differ by more than one lattice site, or are diagonally separated. While Eq.~\eqref{Eq:HubbardHam} formally describes two species of (fermionic) atoms, it can also be straightforwardly adapted to the case of indistinguishable bosons, since bosons and distinguishable fermions have the same underlying $s$-wave interactions in Eq.~\eqref{Eq:3DLowEnergyScatteringAmplitude} and thus the same interaction strength $U$.

The single-particle state $\ket{\vb{j},\sigma,w} \equiv \chd_{\vb{j},\sigma} \ket{0}$ corresponds to the Wannier state in the lowest Bloch band at site $\vb{j}$. This can be obtained from the lowest-band Bloch eigenstates in Eq.~\eqref{eq:BlochStates2}:
\begin{align} \label{Eq:Wannier}
    \ket{\vb{j},\sigma,w} = \frac1{\sqrt{\Omega_D}}\sum_{\kv} e^{i \kv \cdot \vb{j} d}  \ket{\kv,\nuv=\zerov,\sigma},
\end{align}
where $\Omega_D=(2\pi/d)^D$ is the area of the first Brillouin zone, and we have  $\ket{\kv,\nuv,\sv,\sigma}\equiv \ket{\kv,\nuv,\sigma}\otimes\ket{\sv}$, since the different dimensions are separable and the transverse harmonic confinement is spin independent.
Due to the symmetry of the single-particle amplitudes discussed below Eq.~\eqref{Eq:ExpansionEqAtoms}, 
the Bloch wave functions satisfy $\langle \xv \ket{\kv,\nuv=\zerov,\sigma}^* = \langle \xv \ket{-\kv,\nuv=\zerov,\sigma}$, with $\xv$ a position vector along the dimensions of the lattice. 
Equation~\eqref{Eq:Wannier} thus implies
that the Wannier orbitals $w_\sigma (\xv,\vb{j})\equiv \braket{\xv}{\vb{j},\sigma,w}$ are real, and hence that 
the Wannier orbitals are 
maximally localized~\cite{kohn1959}.

We obtain the Hubbard model parameters by projecting onto the lowest-band Wannier states. Specifically, for neighboring sites $\vb{i}$ and $\vb{j}$ along a particular dimension of the lattice, the hopping is given by
\begin{align}
    t^{\vb{i}\vb{j}}_\sigma & =  
    -\frac1{\Omega_1}\int_{-\pi/d}^{\pi/d} dk\, \cos(k d)\, E_{k0}^\sigma,
\end{align}
where $E_{k0}^\sigma$ is the dispersion along the corresponding direction. For the interaction strength $U$, the procedure is more complicated, as we discuss below. To make the connection to the conventional perturbative approach clear, we first consider weak interactions before presenting our $T$-matrix formulation which applies equally well for weak and strong interactions.

\subsubsection{Perturbative approach to calculating Hubbard $U$}
In the standard mapping between the exact and Hubbard Hamiltonians, valid for weak interactions,
the on-site interaction is obtained by assuming a contact interaction of strength $u$, and evaluating
\begin{align} \label{Eq:WannierApprox}
    U \simeq u\int d^D x \,|w_\uparrow(\xv,\vb{j})|^2| w_\downarrow (\xv,\vb{j})|^2,
\end{align}
where $u$ is one of the low-dimensional coupling constants $u_{\rm 1D}$ or $u_{\rm 2D}$ appearing in Eq.~\eqref{eq:u1dandu2d}.
The approximation in Eq.~\eqref{Eq:WannierApprox} is equivalent to first-order perturbation theory 
in $u$. Using Eq.~\eqref{eq:u1dandu2d}, in quasi-1D we find
\begin{align}
    U \simeq -\frac{2}{ma_{\rm 1D}} \int dx \,|w_\uparrow(x,j)|^2| w_\downarrow (x,j)|^2,
    \label{eq:U1Dpert}
\end{align}
while in quasi-2D, up to logarithmic accuracy, we find
\begin{align} \label{Eq:U2DHubbard}
    U \simeq - \frac{2 \pi}{m \log (a_{\rm 2D}/l)} \int d^2x \, |w_\uparrow(\xv,\vb{j})|^2| w_\downarrow (\xv,\vb{j})|^2.
\end{align}
These expressions can now be used together with the definitions of $a_{\rm 1D}$ and $a_{\rm 2D}$ in Eq.~\eqref{eq:a1da2d} to obtain the Hubbard $U$ in terms of the scattering length and parameters of the external potentials. Note that in Eq.~\eqref{Eq:U2DHubbard}, we have ignored the effects of renormalization by assuming $\abs{\log(a_{\rm 2D}/l )/\log(l\Lambda)}\gg1 $.

Our perturbative approach outlined above differs slightly from the commonly quoted approximation~\cite{jakschColdBosonicAtoms1998}
\begin{align} \label{Eq:OldHubbardUDefn}
    U \simeq \frac{4 \pi a}{m} \left( \frac{1}{\sqrt{2\pi} l} \right)^{3-D} \int d^D x \,|w_\uparrow(\xv,\vb{j})|^2| w_\downarrow (\xv,\vb{j})|^2.
\end{align}
This expression ignores the effects of renormalization (i.e., by setting $u\propto \frac{4 \pi a}{m}$) and assumes that the interaction is restricted to the lowest HO level (which, when integrated out, yields $(\sqrt{2\pi} l)^{-1}$ for each dimension of confinement).
By contrast, the perturbative approach in Eqs.~\eqref{eq:U1Dpert} and \eqref{Eq:U2DHubbard} uses the low-dimensional coupling constant $u_{\rm 1D}$ or $u_{\rm 2D}$ appearing in Eq.~\eqref{eq:u1dandu2d}, which includes finite-range effects and the influence of all virtual transitions into higher-lying HO levels.
In the limit of very weak interactions $|a/l| \ll 1$, both methods agree. 

\subsubsection{Hubbard $U$ beyond the perturbative regime}
In order to extend the on-site interaction to stronger interactions, we can instead determine $U$ from equating the exact and Hubbard scattering amplitudes, as was originally done for a 3D cubic lattice~\cite{buchlerMicroscopicDerivationHubbard2010}. That is, we can determine $U$ by requiring
\begin{align} \label{Eq:EquateExactAndHubbardf}
    f(\p)
    = f_H(\p),
\end{align}
for $|\p|\ll 1/d$. Here, $f_H(\p)$ is the scattering amplitude at relative momentum $\p$, as calculated within the Hubbard model \eqref{Eq:HubbardHam}.
Unlike in 3D, we must use a limiting procedure since the scattering amplitudes approach zero in both 1D and 2D as the relative momentum $\p\to \zerov$.
This process of equating scattering amplitudes enforces that the Hubbard model reproduces the exact low-energy scattering wave function.

The Hubbard scattering amplitude can be calculated from the $T$ matrix which, at relative momentum $\pv$, is given by 
\begin{align}
    T^{-1}_H(\p) = \frac{1}{U} - \sum_{\kv \in \rm {1BZ}} \frac{1}{\epsilon^H_{\p} - \epsilon^H_{\kv}+i0},
\end{align}
where 1BZ is the first Brillouin zone. In 1D the dispersion is given by
\begin{align} \label{eq:Hubbard-ek}
    \epsilon^H_{p_z} = -2 (t^z_\uparrow+t^z_\downarrow) \cos(p_z d),
\end{align}
while in 2D it is $\epsilon^H_{\pv} = \epsilon^H_{p_x} + \epsilon^H_{p_y}$. The scattering amplitude for small relative momenta ($|\p| \ll \pi/d$) is related to the $T$ matrix via the same relations as provided in the exact scenario in Eq.~\eqref{Eq:ScatFromT}, with $m_{\rm eff} \to m^H_{\rm eff}$ (i.e., a Hubbard effective mass).

We calculate the interaction $U$ by equating the real parts of (the inverse of) Eq.~\eqref{Eq:EquateExactAndHubbardf}, i.e.,
\begin{align} \label{Eq:EffUDetermine}
    \frac{1}{U} = \lim_{\pv \to \zerov} \left(\frac{m_{\rm eff}^H}{m_{\rm eff}} \Re T^{-1}(\pv) +\mathscr{P}\sum_{\kv \in \rm {1BZ}}  \frac{1}{\epsilon^H_{\p} - \epsilon^H_{\kv}}\right).
\end{align}
Here, we have used the Sokhotski-Plemelj theorem
\begin{align}
    \frac{1}{\alpha+i0} = \mathscr{P} \frac{1}{\alpha} -i \pi \delta(\alpha),
\end{align}
where $\mathscr{P}$ denotes the Cauchy principal value and $\delta$ is the Dirac delta function.

In Eq.~\eqref{Eq:EffUDetermine}, significant care must be taken in quasi-2D since the real parts of 
both terms in the parentheses diverge logarithmically in the limit $\pv \to \zerov$. We can prevent this divergence from creating a computationally unstable problem by considering the imaginary part of (the inverse of) Eq.~\eqref{Eq:EquateExactAndHubbardf} which yields a relationship for the effective masses
\begin{align} \label{Eq:EffMassDetermine}
    \frac{m_{\rm eff}^H}{m_{\rm eff}}= \lim_{\pv \to \zerov} \frac{
    \pi \sum_{\kv \in \rm {1BZ}} \delta(\epsilon^H_{\pv} - \epsilon^H_{\kv}) }{\Im T^{-1}(\pv)}.
\end{align}
This is completely equivalent to the procedure outlined above for the calculation of the effective masses, as explicitly demonstrated in Appendix~\ref{Appendix:DetEffectiveMass}.
In practice, we choose a small $\pv$ to first determine the effective mass ratio using Eq.~\eqref{Eq:EffMassDetermine}, and then, using the same value of $\pv$, we calculate $U$ from Eq.~\eqref{Eq:EffUDetermine}. This procedure ensures that the divergences exactly cancel. Furthermore, it guarantees that the resulting Hubbard $U$ is independent of the precise choice of $m'_{\rm eff}$ in the case of an anisotropic lattice.

\begin{figure*}
    \centering
     \includegraphics[width=\linewidth]{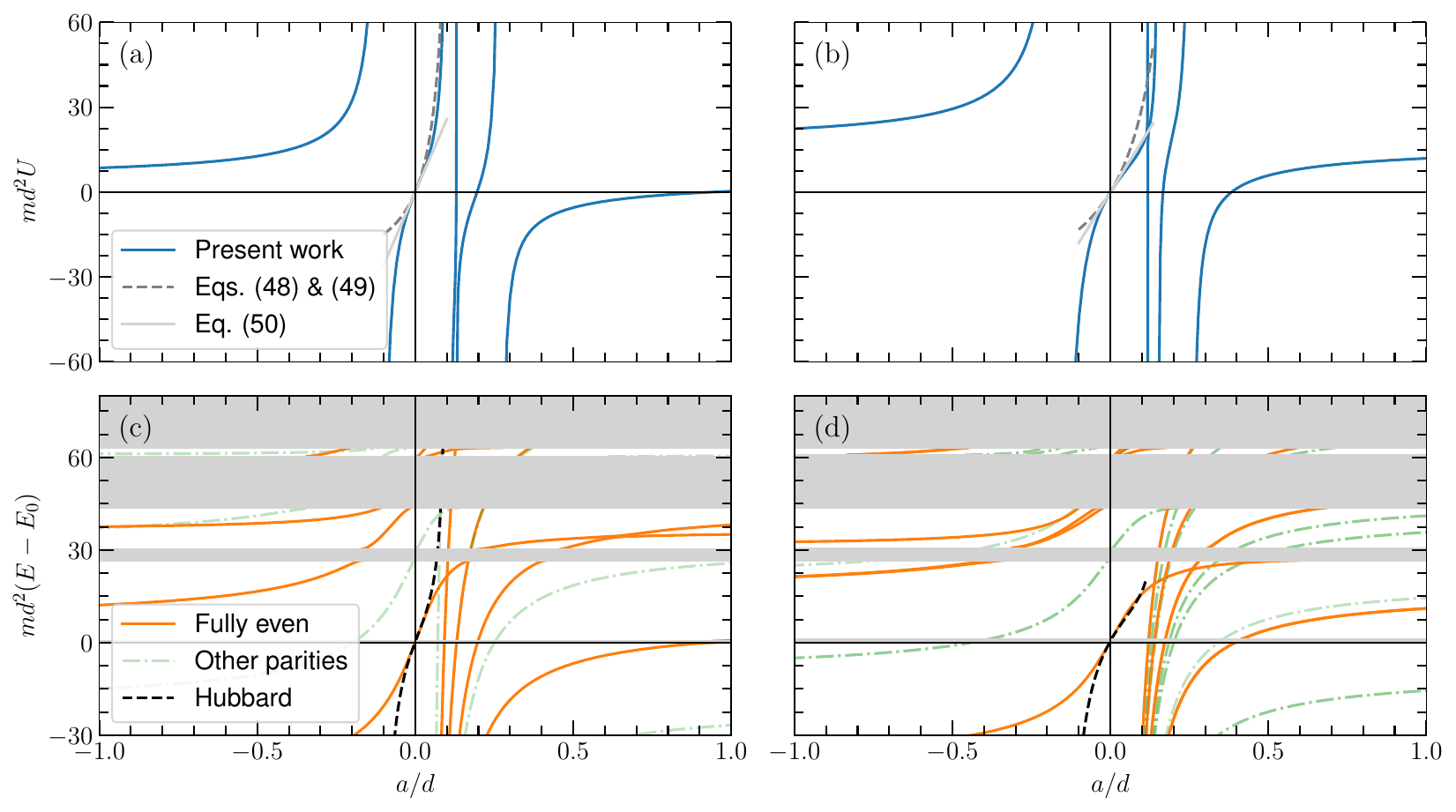}
    \caption{(a,b) Effective on-site interaction $U$ and (c,d) two-body spectrum as a function of the 3D scattering length $a$. We show these for (a,c) quasi-1D and (b,d) quasi-2D lattices, where in both cases $v^i_\sigma = 12 V_r$, $l \simeq 0.13 d$, $R^*=0$, and we consider zero CM quasimomentum. (a,b) The on-site interaction (blue line) is compared with the predictions of perturbation theory (gray dashed line), i.e., Eq.~\eqref{eq:U1Dpert} and~\eqref{Eq:U2DHubbard} for quasi-1D and quasi-2D, respectively, along with the commonly used expression in Eq.~\eqref{Eq:OldHubbardUDefn} (gray, dashed line). (c,d) In the spectra, from bottom to top, we show the first three Bloch bands as well as the continuum (colored gray). We also show the two-body bound-state energies where we color the fully even (other) parity bound states solid orange (dot-dashed green). The bound-state energies are compared alongside those predicted by the Hubbard model (black, dashed line), where $U$ is determined by its corresponding value in (a,b).  \new{Note that the odd parity bound states 
    do not couple to the lowest Bloch band and are thus independent of the interaction $U$}.
    In the spectra, all energies are measured relative to the threshold energy of the lattice $E_0 \equiv E_{\zerov,\zerov;\zerov,\zerov;0}$. For clarity, we only plot the first ten two-body bound states in panel (c) and the first 20 in panel (d). Note that the lowest energy state is always even parity and is outside the range plotted in (d).}
    \label{fig:UandBoundStates}
\end{figure*}

\begin{figure}
    \centering
    \includegraphics[width=\linewidth]{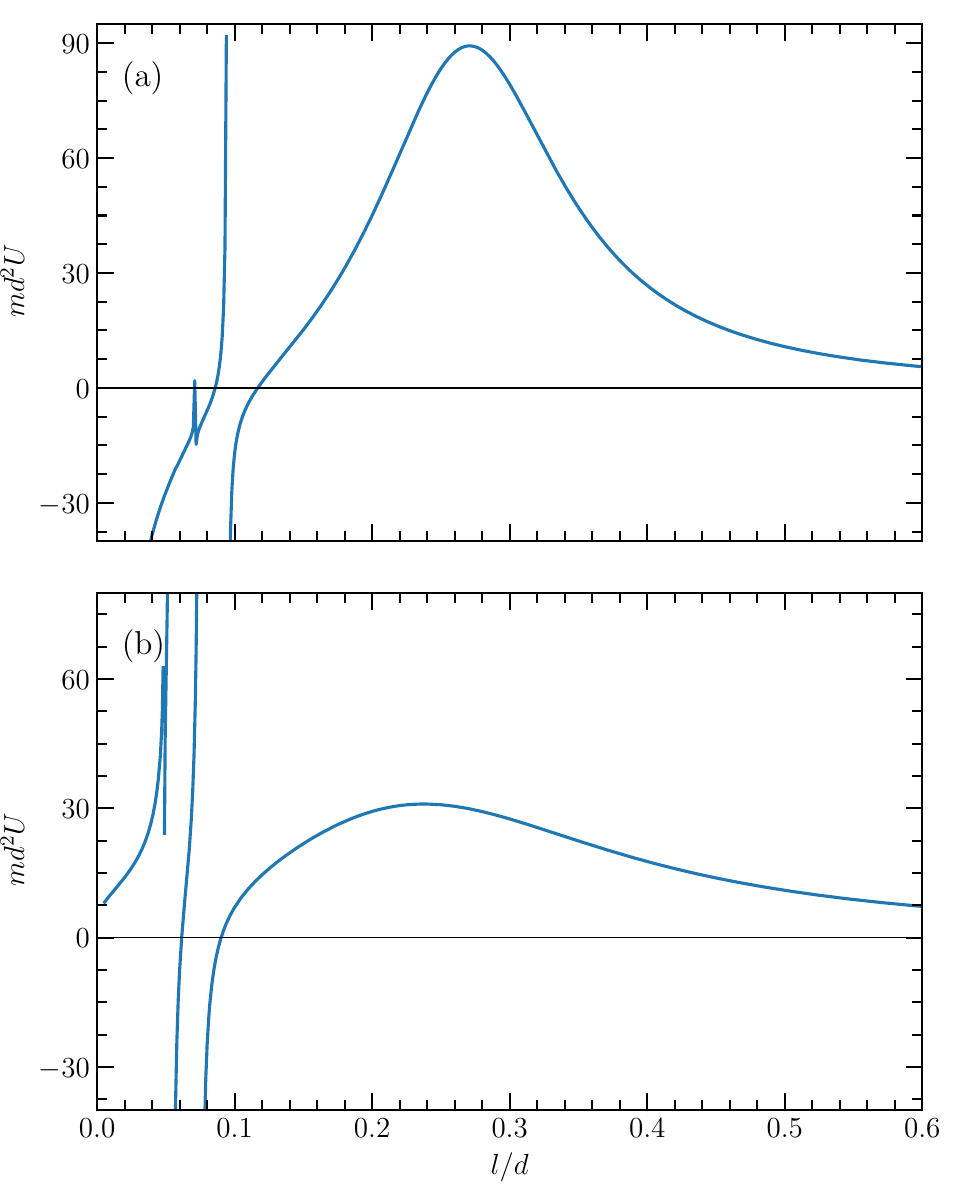}
    \caption{The on-site interaction $U$ at unitarity (i.e., $1/a=0$) as a function of the harmonic oscillator length in (a) quasi-1D and (b) quasi-2D. Remarkably, even at unitarity the lattice can induce resonances in $U$, which can be extremely narrow as seen for $l/d \sim 0.05$. In both cases we consider a broad Feshbach resonance ($R^*=0$) and take $v^i_\sigma = 12V_r$.}
    \label{fig:ResonancesUnitarity}
\end{figure}

\begin{figure}
    \centering
    \includegraphics[width=\linewidth]{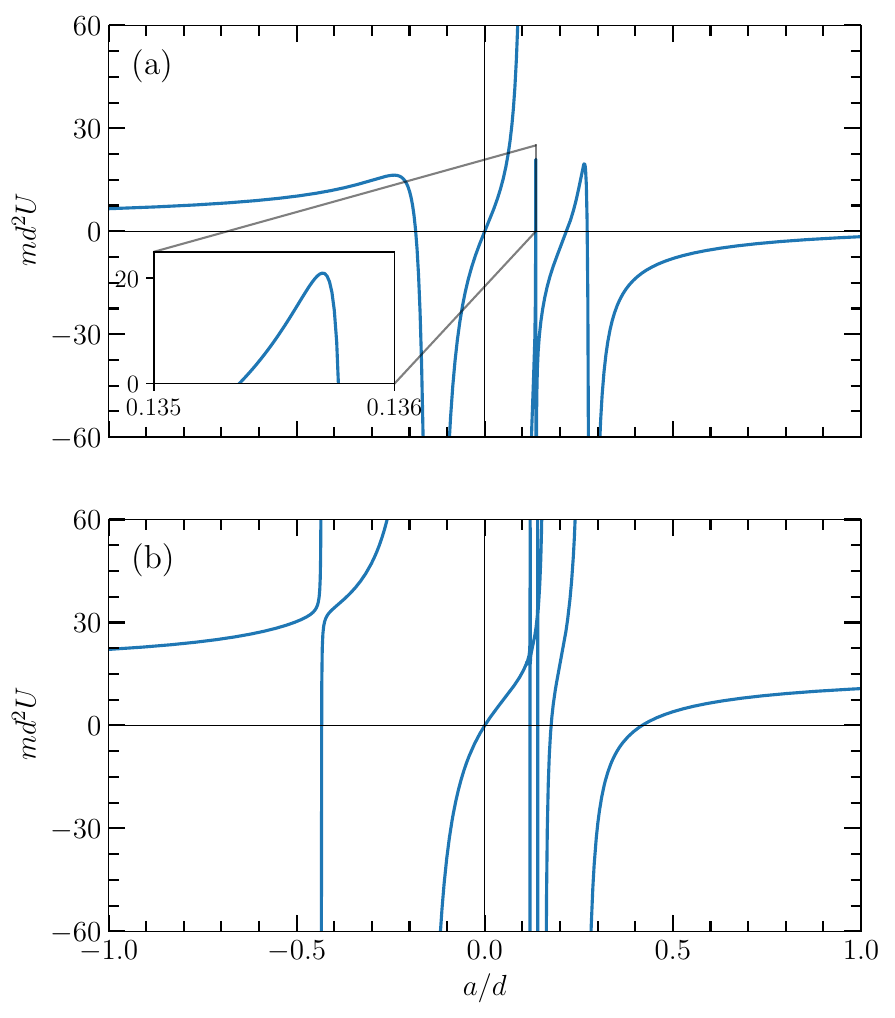}
    \caption{The on-site interaction $U$ as a function of 3D scattering length $a$ for a state-dependent lattice in (a) quasi-1D and (b) quasi-2D, where $v^i_\uparrow = 12V_r$, $v^i_\downarrow=10V_r$, $l\simeq 0.13d$ and $R^*=0$. The state-dependent lattice displays highly narrow resonances, which peak at both finite and infinite values. The inset in panel (a) shows a zoomed in region of a highly narrow, finite resonance.}
    \label{fig:StateDependentLatticeU}
\end{figure}

\subsection{Numerical results} \label{Sect:DetermineHub}
To investigate the behavior of the on-site Hubbard interaction, we consider zero CM quasimomentum ($\qv=\zerov$) and focus on the limit of a broad Feshbach resonance where \new{$R^*$ is small compared to all other length scales such that we can take $R^*\to0$.}
It is instructive to begin with the simplest scenario of a lattice with no spin or directional dependence ($v = v^{i}_\sigma$). Figure~\ref{fig:UandBoundStates}(a,b) shows our calculated Hubbard interaction $U$ as a function of the scattering length $a$ for both quasi-1D and quasi-2D, where we set the strength of the optical lattice $v =12 V_r$ ($t = t^{\vb{i}\vb{j}}_\sigma \simeq 0.06 md^2$) and the harmonic oscillator length $l\simeq 0.13 d$ (corresponding to strong confinement).  We see that the exact calculation of $U$  shows a rich behavior with increasing 
$a/d$. 
In particular, we find multiple broad resonances for sufficiently large $|a|$ which can, in principle, be used to tune the Hubbard interaction, similarly to how the 3D scattering length is tuned close to a Feshbach resonance.

In the weak-coupling limit $|a|/d \ll 1$, 
our calculated Hubbard $U$ matches well with the perturbative expressions in Eqs.~(\ref{eq:U1Dpert}-\ref{Eq:OldHubbardUDefn}). 
Here it appears that our new expressions based on the low-dimensional scattering lengths typically diverge faster from the exact result than the standard approximation~\eqref{Eq:OldHubbardUDefn}.  
However, in the quasi-1D geometry, we find that 
Eq.~\eqref{eq:U1Dpert} captures the appearance of the first broad resonance at $a \approx 0.1 d \approx l$, indicating that this is due to a resonance in $a_{\rm 1D}$, i.e., 
it is directly linked to the underlying confinement-induced resonance 
in the absence of a lattice~\cite{olshaniiAtomicScatteringPresence1998}. 
All the other resonances in both quasi-1D and quasi-2D geometries are induced by the lattice, although, as discussed below, they resemble the quasi-1D confinement-induced resonance since they arise from higher energy bands~\cite{Bergeman2003}.
Similar lattice-induced resonances have previously been investigated both theoretically~\cite{vonstecher2011} and experimentally~\cite{rieggerLocalizedMagneticMoments2018} for 1D and quasi-1D lattice geometries, respectively.

The behavior of the Hubbard $U$ is intimately connected to the two-body spectrum, shown in Fig.~\ref{fig:UandBoundStates}(c,d). \new{In addition to the Bloch bands that originate from the single-particle dispersions,}
the spectrum consists of both attractive and repulsive bound states (below and above the lowest Bloch band, respectively), obtained from our full numerical calculation. 
In particular, we see that the zeros of $U$ correspond to the points at which bound states of fully even parity merge into the lowest Bloch band from below. \new{This behavior is qualitatively different from the situation in 3D, where the crossing of a bound state into the lowest Bloch band typically leads to a scattering resonance~\cite{buchlerMicroscopicDerivationHubbard2010}. 
Instead, we find that the even-parity bound state remains below the lowest band at the points where $U$ diverges, which is reminiscent of the situation for a confinement-induced resonance in a quasi-1D uniform system~\cite{Bergeman2003}.} 
\new{Finally, we stress that all states of other parities are completely independent of the low-energy scattering parameterized by $U$ since they}
are decoupled from the lowest Bloch band, i.e., ${\mathcal H}^{\zerov ,\nuv}_{\kv,\zerov,\zerov} = 0$ when any component of $\nuv$ is odd (see the discussion of symmetries in Sec.~\ref{Sect:OpenChannelTMatrix}). 
\new{This again resembles the quasi-1D uniform case, where bound states involving different centre-of-mass HO quantum numbers are similarly decoupled from the scattering properties.}

We also compare our results for the two-body \new{even-parity} bound states with those obtained within the Hubbard model, where the bound-state energies $E_{2b}$ satisfy
\begin{align}
    \frac{1}{U} - \sum_{\kv \in \rm {1BZ}} \frac{1}{E_{2b} - \epsilon^H_{\kv}} = 0 \, ,
\end{align}
with the Hubbard dispersion $\epsilon^H_{\kv}$ as defined in Eq.~\eqref{eq:Hubbard-ek}. In the regime $|U|/t\gg 1$, this gives $E_{2b} \approx U$ for the attractive and repulsive bound states in the lowest band. 
We compare this approximate Hubbard result with the exact energies in Fig.~\ref{fig:UandBoundStates}(c,d), and we see that this only describes the bound-state energies when $|a/d| \ll 1$.
Indeed, the Hubbard model is unable to reproduce the exact bound state once $|U|$ 
is comparable to the bandgap between lowest Bloch bands, 
similarly to the case in a 3D lattice~\cite{buchlerMicroscopicDerivationHubbard2010}. 
Thus, enforcing that $|U|$ is much smaller than the bandgap in the case of a deep, tightly confined lattice,  
we obtain the followings condition for the validity of the Hubbard model in describing bound states
\begin{subequations}
\begin{align}
    \left|\frac{a}{d} \right| &\ll \left( \frac{v}{V_r} \right)^{1/4} \sqrt{\frac{2 V_r}{\pi \omega}}, &&\rm 1D,\\
    \left|\frac{a}{d}\right| &\ll \sqrt{\frac{V_r}{\pi \omega}}, &&\rm 2D.
\end{align}
\end{subequations}
For comparison, the corresponding condition in 3D is $|a/d| \ll (V_r/v)^{1/4}/2 \sqrt{\pi}$~\cite{jakschColdBosonicAtoms1998,buchlerMicroscopicDerivationHubbard2010}. 

While our effective Hubbard $U$ cannot capture the two-body bound states for arbitrary scattering length $a$, we expect it to provide an accurate description of the low-energy (unbound) scattering properties. Thus, the Hubbard model can still be used to describe repulsive many-body ground states.  
Indeed, this situation is similar to quasi-1D and quasi-2D gases in the absence of a lattice, where the many-body physics of interest can be dominated by scattering states that are well captured by a simple 1D or 2D model, even though the bound states are no longer simply parameterized by an effective 1D or 2D scattering length~\cite{levinsenStonglyInteractingTwo2015,Levinsen2012,Bergeman2003,paredesTonksGirardeauGas2004}.
\new{Furthermore, the presence of low-energy bound states with other parities (Fig.~\ref{fig:UandBoundStates}) does not pose an issue for the low-energy description of scattering in the lowest Bloch band since these are fully decoupled; this is again similar to quasi-1D and quasi-2D uniform systems where there are many zero-energy crossings of center-of-mass excitations of the two-body bound states that can safely be neglected for the purposes of defining an effective low-energy interaction strength since they involve harmonic oscillator quantum numbers that are decoupled from the lowest level~\cite{buschTwoColdAtoms1998}.}

We also examine $U$ at unitarity ($1/a=0$), where the standard perturbative expression~\eqref{Eq:OldHubbardUDefn} completely fails. From Fig.~\ref{fig:UandBoundStates}, we see that $U$ saturates to a finite (repulsive) value as $a \to \pm\infty$ for the particular ratio $l/d \simeq 0.13$. However, once we vary $l/d$, we find that resonances in $U$ exist for $l/d \lesssim 0.1$, as shown in Fig.~\ref{fig:ResonancesUnitarity}.
This is particularly remarkable given that the 1D and 2D scattering lengths in Eq.~\eqref{eq:a1da2d} simply scale as the harmonic oscillator length in this limit. These resonances are therefore purely a consequence of the non-trivial impact on the interactions due to the harmonic confinement and the lattice along orthogonal directions.

For the case of a state-dependent lattice, we find even richer behavior for $U$, as shown in Fig.~\ref{fig:StateDependentLatticeU}. 
Here we have used the same parameters as in Fig.~\ref{fig:UandBoundStates}, apart from the lattice depth which we take to be $v^i_\uparrow = 12V_r$ and $v^i_\downarrow = 10 V_r$. 
In a state-dependent lattice, bound states of arbitrary parity are coupled to the lowest Bloch band, which adds significant complexity to the structure of $U$ as a function of $a$. In both quasi-1D and quasi-2D, this structure is evident in the appearance of a large number of narrow resonances relative to Fig.~\ref{fig:UandBoundStates}. Remarkably, in quasi-1D many of these resonances no longer diverge to infinite values, unlike the case in quasi-2D.

\subsection{Comparison to experiment}

\begin{figure}
    \centering
    \includegraphics[width=\linewidth]{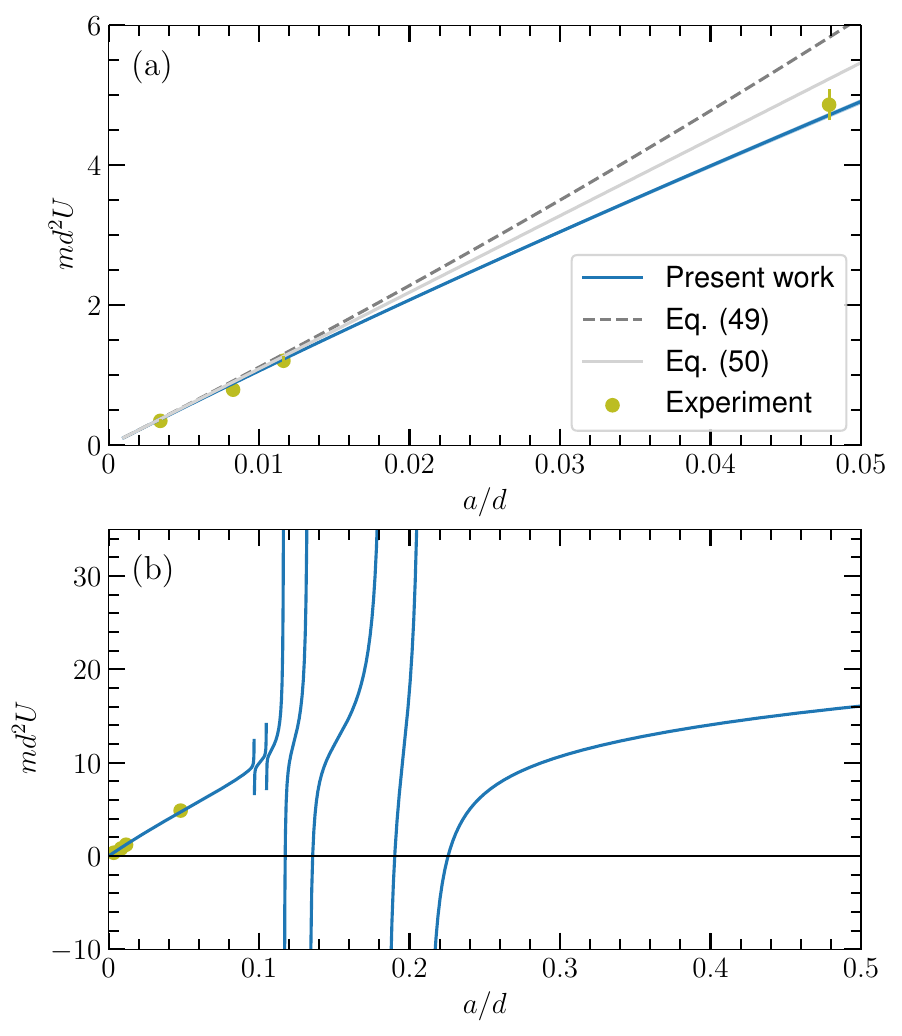}
    \caption{The on-site interaction $U$ as calculated using our theory (blue line) and experimentally extracted using lattice modulation spectroscopy~\cite{greifSiteresolvedImagingFermionic2016} (olive points) for $v^x_\sigma = 12.5 V_r$, $v^y_\sigma = 15.9 V_r$, $\omega \simeq 3.71 V_r$ and $R^*=0$. Panel (a) shows the comparison, alongside perturbation theory, Eq.~\eqref{Eq:OldHubbardUDefn} %
    (gray, dashed line), on an experimentally relevant scale of scattering lengths. Panel (b) extends to larger scattering lengths showing that the first resonance is located at roughly twice the largest scattering length considered in experiment.}
    \label{fig:ComparisonGrief}
\end{figure}

Recent quantum gas experiments often operate in a regime that goes beyond the 
validity of standard perturbative approaches, and thus require the Hubbard $U$ to be determined experimentally through 
the use of lattice modulation spectroscopy~\cite{schaferToolsQuantumSimulation2020}.
Starting from a Mott insulating state, the lattice along one dimension is modulated in depth at variable frequency. 
When the frequency matches the on-site interaction energy 
$U$, the lattice sites can become doubly occupied. This results in an observable reduction in lattice filling, since doubly-occupied lattice sites are detected as empty sites in experiment~\cite{greifSiteresolvedImagingFermionic2016}, thus allowing the experimental determination of $U$.

Figure~\ref{fig:ComparisonGrief} compares our theoretical determination of $U$ with the results of lattice modulation spectroscopy for a quasi-2D gas of fermionic ${}^6$Li atoms in an equal mixture of the two lowest hyperfine ground states in a square lattice~\cite{greifSiteresolvedImagingFermionic2016}.
Crucially, we find that our effective $U$ is in excellent agreement with the experimental value (without any fitting parameters), even though there are noticeable deviations from perturbation theory with increasing $a/d$ [panel (a)]. 
While the experiment goes beyond the validity of 
perturbation theory, the largest scattering length considered is still below that required to access the first resonance in $U$, as shown in panel (b).
However, we stress that the first resonance occurs already at $a \approx 0.1 d$, or equivalently $a \approx 1100 a_B$ (with $a_B$ the Bohr radius), which is experimentally attainable in ${}^6$Li mixtures by virtue of the available magnetic Feshbach resonance.

\new{Finally, we emphasize that our calculation for $U$ is correct as long as the quantum gas is described by the Hubbard model. Therefore, experiments for which the effective low-energy description fails are no longer Hubbard-model simulators; rather, they must be modeled by theories beyond the standard Hubbard description.}

\section{Conclusion} \label{Sect:conclusion}

To conclude, we have exactly solved the problem of two atoms interacting in both a quasi-1D and quasi-2D optical lattice. In particular, beginning with a microscopic model for two atoms interacting in a lattice with transverse harmonic confinement, we have provided a numerically exact calculation of the two-body $T$ matrix. Here we have treated the lattice with significant generality, accounting for a potential dependence on both the internal states of the atoms, as well as an asymmetry in the lattice strengths along the different directions in the quasi-2D case. Using this calculation, we derived the scattering amplitude of two atoms in the lowest Bloch band, which includes all possible virtual transitions into higher Bloch bands and is appropriately renormalized. We have also provided significant detail on making this calculation numerically tractable in the Appendices, and have made the code freely available on GitHub~\cite{SourceCode}.

We have used our exact solution of the two-body problem 
to determine the effective Hubbard on-site interaction $U$, similarly to the work of Ref.~\cite{buchlerMicroscopicDerivationHubbard2010} on the 3D cubic optical lattice. 
\new{This provides the best possible calculation for the Hubbard $U$ when the atoms in a lattice are described by the Hubbard model.}
By varying the scattering length and confinement strength, we have shown that $U$ displays a rich behavior, including many broad resonances. 
Furthermore, we have identified qualitatively different behavior between the quasi-1D and quasi-2D systems of the present work, and the 3D system considered by Ref.~\cite{buchlerMicroscopicDerivationHubbard2010}. 
We have also demonstrated that our results for $U$ agree well with those obtained experimentally for a quasi-2D square lattice in a quantum gas microscope~\cite{greifSiteresolvedImagingFermionic2016}. 
We expect the broad resonances in $U$ to be within reach of current experiments, with the intriguing prospect of realizing strongly correlated phases in the vicinity of such resonances.

In future studies, our formalism can be generalized to multi-band models and 
more exotic geometries, such as triangular lattices, which have recently been imaged with single-site resolution for both bosonic~\cite{yamamotoSinglesiteresolvedImagingUltracold2020} and fermionic atoms~\cite{yangSiteResolvedImagingUltracold2021}. Beyond ultracold atomic gases, extensions of our work also hold promise for the precise characterization of exciton-exciton and exciton-electron interactions in emerging designer lattices, such as moir\'e superlattices in twisted bilayers of atomically thin materials~\cite{Bistritzer2011}.

\section*{Acknowledgments}
We gratefully acknowledge fruitful discussions with Nelson Darkwah Oppong, Dmitry Efimkin and Hans Peter B\"uchler. HSA acknowledges support through an Australian Government Research Training Program Scholarship. JL and MMP acknowledge support from the Australian Research Council Centre of Excellence in Future Low-Energy Electronics Technologies (CE170100039). JL and MMP are also supported through Australian Research Council Future Fellowships FT160100244 and FT200100619, respectively.

\appendix

\section{Two-channel Hamiltonian} \label{Appendix:TwoChannelHam}
Here we provide details on the derivation of the two-channel interaction $\Hh_{\rm co}$ in Eq.~\eqref{Eq:InteractionHamiltonian}. The standard form of the two-channel interaction is~\cite{timmermansFeshbachResonancesAtomic1999}
\begin{align} \label{eq:2channelint}
    \Hh_{\rm co} = & g \sum_{\qvt, \kvt}\chi_{\rm{3D}}(\kvt)\big( \dhd_{\qvt} \ch_{\qvt/2 + \kvt, \downarrow} \ch_{\qvt/2 - \kvt, \uparrow} \nonumber \\ &\qquad\qquad\qquad\qquad + \rm{h.c.} \big),
\end{align}
where the subscripts 
indicate that these are 
3D vectors. 
Focusing for simplicity on quasi-2D where the transverse motion does not need to be regularized, we define $\Kv$ and $K_z$ to be the momentum along and transverse to the lattice, respectively. We then perform a change of basis according to
\begin{align}
    \ch_{\kvt,\sigma} &= \sum_{s} \int dz \, e^{-iK_z z} \phi^A_{s}(z) \, \ch_{\Kv s\sigma}, \label{Eq:QBasisTransform1}\\
    \dh_{\kvt, \sigma} &= \sum_{S} \int dz \, e^{-iK_z z} \phi^M_{S} (z) \, \dh_{\Kv S} \label{Eq:QBasisTransform2}.
\end{align}
Here, $\phi^A_{s}(z)$ $\left(\phi^M_{S}(z)\right)$ 
are the HO eigenstates of the atoms (molecule) with quantum numbers $s$ and $S$, respectively, and we take these to be real. Applying this basis transformation and focussing only on the $z$ part of the momentum sums in Eq.~\eqref{eq:2channelint}, we find 
\begin{align}
 &   \sum_{Q_z K_z} \int dz_1 \,dz_2 \, dz_3\, \phi^M_S(z_1) \phi^A_{s_1}(z_2) \phi^A_{s_2}(z_3) \nonumber\\
    &{}\qquad \times e^{iQ_z(z_1-z_2/2-z_3/2)+iK_z(z_2-z_3)} \nonumber \\
    &=\int dz \, \phi^{M}_S(z) \phi^A_{s_1} (z) \phi^A_{s_2} (z). 
\end{align}
This can be related to the transformation from the CM frame to that of the individual particles as follows:
\begin{align}
\label{eq:deltaint}
&\int dz \, \phi^{M}_S(z) \phi^A_{s_1} (z) \phi^A_{s_2} (z) 
\nonumber \\ =&
\int \frac{dz\,dz_{\rm r}}{\phi_s(0)}
\phi^M_S(z)\phi_s(z_{\rm r})\delta^{(2)}(z_{\rm r})\phi^A_{s_1}(z+z_{\rm r}/2)\phi^A_{s_2}(z-z_{\rm r}/2)
\nonumber \\ \equiv &
\frac1{\phi_s(0)}
\bra{S \,s}\delta^{(2)}(\hat z_{\rm r})\ket{s_1s_2}.
\end{align}
%
Inserting a complete set of harmonic oscillator states of the center of mass and relative motion we find
\begin{align}
&\frac1{\phi_s(0)}
\bra{S \,s}\delta^{(2)}(\hat z_{\rm r})\ket{s_1s_2}
\nonumber \\ =&\frac1{\phi_s(0)}\sum_{S's'}
\bra{S \,s}\delta^{(2)}(\hat z_{\rm r})\ket{S's'}\braket{S's'}{s_1s_2}\nonumber \\
=&\sum_{s'}\phi_{s'}(0)\braket{Ss'}{s_1s_2}.
\end{align}
Here we have used the fact that the interaction is decoupled from the center of mass motion in the transverse direction. Gathering terms then yields Eq.~\eqref{Eq:InteractionHamiltonian}, as given in the main text. The same argument straightforwardly generalizes to the case of quasi-1D. 

It is now straightforward to obtain the matrix element 
of the interaction. In the absence of a lattice, we obtain Eq.~\eqref{eq:intmatrixelt} via
\begin{align} \label{eq:intmatrixeltapp}
\nonumber
   & \bra{0} \dh_{\zerov\zerov} \Hh_{\rm int} \ket{\Kv',\sv'} \\ & =
    \bra{0} \dh_{\zerov\zerov}   \sum_{\substack{\Kv \Qv \\ \sv_1 \sv_2 \Sv}} \xi^{\Sv \Kv}_{\sv_1 \sv_2} \dhd_{\Qv \Sv} \ch_{\Qv/2+\Kv,\sv_2\downarrow} \ch_{\Qv/2-\Kv,\sv_1\uparrow} \nonumber \\ &\qquad \times \sum_{\sv_1',\sv_2'}\braket{\sv_1',\sv_2'}{\Sv'=0,\sv'}
    \chd_{-\Kv', \sv_1', \uparrow}  \chd_{\Kv', \sv_2', \downarrow} \ket{0}\nonumber \\
    &=\sum_{\sv_1 \sv_2 \sv}
    \chi(s,K') \, \phi_{\sv}
\braket{\Sv=0,\sv}{\sv_1,\sv_2}\braket{\sv_1,\sv_2}{\Sv'=0,\sv'}
    \nonumber \\
    &=
    \phi_{\sv'} \chi(s', K').\nonumber
\end{align}

In the presence of a lattice, we instead calculate Eq.~\eqref{Eq:HMatElementsExact} by taking advantage of the separability of the interaction. Specifically, we have
\begin{align}
    &\bra{\qv, \nuv} \Hh_{\rm int} \ket{\kv, \qv; \nuv_1, \nuv_2; \sv}\nonumber\\
    &= \phi_{\sv}\chi_1(s)\sum_{\Kv\Qv\nv_1\nv_2\Nv}\chi_2(K)\eta_{\Nv}^{(\qv,\nuv)}\varphi_{\nv_1}^{(\qv/2-\kv,\nuv_1,\uparrow)}\varphi_{\nv_2}^{(\qv/2+\kv,\nuv_2,\downarrow)}
    \nonumber \\ & \times \delta_{\Qv,\qv+2\pi \Nv/d}\delta_{\Qv/2+\Kv,\qv/2+\kv+2\pi \nv_2/d}\delta_{\Qv/2-\Kv,\qv/2-\kv+2\pi \nv_1/d}\nonumber \\
    &=\phi_{\sv}\chi_1(s)\sum_{\Nv,\nv} 
    \chi_2(2 \pi |\nv|/d)
    \eta^{(\qv,\nuv)}_{\Nv} \varphi^{(\qv/2-\kv,\nuv_1,\uparrow)}_{\Nv/2+\nv} \varphi^{(\qv/2+\kv,\nuv_2,\downarrow)}_{\Nv/2-\nv}.
\end{align}
In evaluating the sum over Kronecker delta functions, we have defined $\nv=\Nv/2+\nv_1$, where the sums over $\Nv$ and $\nv_1$ run over integers. Therefore, the sum over $\nv$ runs over integers when $\Nv$ is even and over half-integers when $\Nv$ is odd.

 \section{Numerical implementation} \label{Appendix:NumImplement}
While the equations provided in Sec.~\ref{Sect:OpenChannelTMatrix} for the calculation of the open-channel $T$ matrix are exact, they remain difficult to calculate numerically. In particular, proper regularization of the contact interactions requires the numerical grids to be truncated carefully; truncating the number of atomic Bloch bands, for example, has been shown to lead to a systematic error~\cite{buchlerMicroscopicDerivationHubbard2010}. We showed in Section~\ref{Sect:Renormalization} that in the absence of the lattice, the contact interactions can be regularized with an ultraviolet cutoff that acts on the relative motion of the atoms. This remains true when the lattice is included. The challenge is that the matrix elements of the polarization bubble $\hat \Pi$ in Eq.~\eqref{Eq:PiElements} are written in terms of the individual Bloch states. To circumvent this, 
we evaluate the matrix elements of $\hat \Pi$ using an atom-atom basis given by
\begin{align} \label{Eq:atom-atombasis}
    \ket{\kv, \qv; \Nv, \nv;\sv}  &\equiv \sum_{\sv_1,\sv_2} 
    \braket{\sv_1,\sv_2}{\Sv=0,\sv}
    \nonumber \\ &\hspace{-20mm}\times
    \ket{\qv/2-\kv; \Nv/2+\nv, \sv_1,\uparrow} 
    \otimes \ket{\qv/2+\kv; \Nv/2-\nv,\sv_2,\downarrow},
\end{align}
where
\begin{align}
    \ket{\kv; \nv,\sv ,\sigma} \equiv \chd_{\kv + 2\pi \nv/d, \sv, \sigma} \ket{0}.
\end{align}
We remind the reader that the regularization of the contact interactions amounts to cutting off the sum over $\sv$ in quasi-1D, and the sum over $\nv$ in quasi-2D [see Eq.~\eqref{eq:OurRenormEq}].

In calculating the polarization bubble, we must first use the atom-atom basis in Eq.~\eqref{Eq:atom-atombasis} to diagonalize the 
non-interacting part of the Hamiltonian, $\hat H_\uparrow + \hat H_\downarrow$. We find that in order to obtain convergence of the polarization bubble matrix, we require only a small number of CM momentum states $\Nv$, but a large number of $\nv$ states (since the contact interactions couple low and high momentum states with a constant coefficient). However, we can drammatically speed up the calculation by taking advantage of the fact that the lattice is irrelevant at high energies. In particular, for
$n_r \gg d \sqrt{m v^i_\sigma}/2 \pi$ and $|n| > n_r$, Eq.~\eqref{Eq:ExpansionEqAtoms} reduces to that of a free atom:
\begin{align} \label{Eq:ApproxFreeParticle}
    E^{\sigma}_{k_i \nu_i} \varphi_n^{(k_i,\nu_i,\sigma)} &\simeq  \epsilon_{k_i,n} \varphi_n^{(k_i,\nu_i,\sigma)}.
\end{align}
Therefore, for each component of $\nv$ satisfying $|n_i| > n_r$, the atom-atom basis in Eq.~\eqref{Eq:atom-atombasis} approximately diagonalizes the optical lattice, i.e.,
\begin{align}
    (\hat H_{\uparrow} + \hat H_{\downarrow}) \ket{\kv, \qv; \Nv, \nv;\sv} \simeq \epsilon_{\kv,\qv;\Nv,\nv;2s} \ket{\kv, \qv; \Nv, \nv;\sv},
\end{align}
where $\epsilon_{\kv,\qv;\Nv,\nv;2s} \equiv \epsilon_{\qv/2-\kv,\Nv/2+\nv} + \epsilon_{\qv/2+\kv,\Nv/2-\nv} + 2s\omega$. This enables us to separate the low and high energy (relative to the lattice energy)
components of $\Pi$ in Eq.~\eqref{Eq:PiElements} according to 
\begin{align} \label{Eq:PolBubbleLandH}
    &\Pi^{\qv}_{\nuv \nuv'}
    =\sum_{\kv\nuv_1\nuv_2s}
    |\phi_{2s}|^2 \mathcal{H}^{\qv,\nuv}_{\kv \nuv_1 \nuv_2} \frac{
    \chi_1^2(2s)}{E - E_{\kv, \qv; \nuv_1, \nuv_2;2s}}
    \mathcal{H}^{\qv,\nuv'}_{\kv \nuv_1 \nuv_2} \nonumber \\ &{}\quad
    \simeq \sum_{\kv \nuv_1 \nuv_2 s} |\phi_{2s}|^2 \tilde{ \mathcal{H}}^{\qv,\nuv}_{\kv \nuv_1 \nuv_2} \frac{\chi_1^2(2s)}{E - E_{\kv, \qv; \nuv_1, \nuv_2;2s}}
    \tilde{ \mathcal{H}}^{\qv,\nuv'}_{\kv \nuv_1 \nuv_2} \nonumber\\
    &{}\qquad +\sum_{\kv \Nv \nv  s}^>  |\phi_{2s}|^2 \eta^{(\qv,\nuv)}_{\Nv} \frac{\chi_1^2(2s)\chi_2^2( 
    2\pi |\nv|/d)}{E - \epsilon_{\kv,\qv;\Nv,\nv;2s}}\eta^{(\qv,\nuv')}_{\Nv} \nonumber \\
    &{} \quad\equiv \Pi^{\qv (\rm LE)}_{\nuv \nuv'} + \Pi^{\qv(\rm HE)}_{\nuv \nuv'},
\end{align}
where 
the low energy matrix elements
\begin{align}
    \tilde{\mathcal{H}}^{\qv \nuv}_{\kv \nuv_1\nuv_2} &= \sum_{\Nv \nv}^< \bra{\qv,\nuv} \Hh^L_{\rm int} \ket{\kv,\qv; \Nv,\nv} \nonumber\\
    &{} \hspace{5em} \times\langle{\kv,\qv; \Nv,\nv}| {\kv,\qv ; \nuv_1,\nuv_2} \rangle.
\end{align}
Here, the $<$ ($>$) notation indicates that the sum over $\nv$ or $\nv'$ is restricted such that the components satisfy $|n_i|$, $|n_i'|\leq n_r$ ($|n_i|>n_r$). We can view $\Pi^{(\rm HE)}$ as the asymptotic high energy correction to the polarization bubble. This correction term can be computed efficiently using standard numerical integration techniques.

\section{Calculation of on-shell \textit{T} matrix} \label{Appendix:OnShellTMat}
The calculation of the retarded on-shell $T$ matrix requires the analytic continuation of the $T$ matrix onto the real energy axis (from the upper-half complex plane).
Since we require the $T$ matrix for energies in the first Bloch band, this means that we must take care in calculating the $\nuv_1=\nuv_2=\zerov$, $s=0$ contribution to the polarization bubble, i.e.,
\begin{align} \label{Eq:polBubblePoleCont} 
    {\Pi}^{\zerov}_{\nuv \nuv'} \equiv \sum_{\kv} |\phi_{0}|^2 \mathcal{H}^{\zerov,\nuv}_{\kv \zerov \zerov} \frac{1}{E - E_{\kv}+i0}
    \mathcal{H}^{\zerov,\nuv'}_{\kv \zerov \zerov },
\end{align}
where we now focus on zero CM quasimomentum (i.e. $\qv=\zerov$). We remind the reader that $E_{\kv}$ is the energy of the two atoms in the lowest band at relative momentum $\kv$ (and $\qv=0$), see Sec.~\ref{subsub:scat}. In the following, we perform the integral over $\kv$ using the Sokhotski Plemelj theorem
\begin{align}
    \frac{1}{E - E_{\kv}+i0} = \mathscr{P} \frac{1}{E - E_{\kv}} -i \pi \delta(E - E_{\kv}),
\end{align}
where $\mathscr{P}$ indicates the principal part.

In quasi-1D this approach is simple to implement. At each energy $E$, we find the position of the pole in Eq.~\eqref{Eq:polBubblePoleCont} by solving $E - E_{k}=0$ for $k$. This yields two solutions which we denote by $k=\pm k_p$. The principal value component is then determined by integrating evenly around the pole positions. For example, to integrate around the pole $k=k_p$, we consider the region $[k_p-\Delta,k_p+\Delta]$ (for some $\Delta>0$). Using the definition of the principal value, we have
\begin{align} \label{Eq:PrincipaValue}
    \mathscr P \int_{k_p-\Delta}^{k_p+\Delta} \frac{dk}{E - E_{k}} =  \lim_{\alpha \to 0} \left( \int_{k_p-\Delta}^{k_p- \alpha}+\int_{k_p+\alpha}^{k_p+\Delta} \right) \frac{dk}{E - E_{k}}.
\end{align}
To evaluate these integrals we employ $N$-point Gauss-Legendre quadrature, which defines nodes $k^{}_i$ and weights $w^{}_i$ such that
\begin{align} \label{Eq:Int1}
    \int_{k_p-\Delta}^{k_p} \frac{dk}{E - E_{k}} \approx \sum_{i=1}^N \frac{w^{}_i}{E - E_{k^{}_i}}.
\end{align}
Similarly, the same nodes and weights can be used to approximate the second integral via
\begin{align} \label{Eq:Int2}
    \int_{k_p}^{k_p+\Delta} \frac{dk}{E - E_{k}} \approx \sum_{i=1}^N \frac{w_{N+1-i}}{E - E_{k^{}_{N+1-i}+\Delta}},
\end{align}
where we have reversed the nodes and shifted them by $\Delta$, while also reversing the weights. Combining Eqs.~\eqref{Eq:Int1} and~\eqref{Eq:Int2}, then yields the principal value. By reversing the nodes and weights in this fashion, we always integrate evenly around the pole at $k=k_p$. Moreover, owing to the fact that the Gauss-Legendre quadrature never sets a node equal to the end-point of the integration, we can effectively ignore $\alpha$ in Eq.~\eqref{Eq:PrincipaValue} (increasing using a higher $N$-point quadrature rule will take the limit of $\alpha \to 0$). Finally, the
Dirac delta contribution is calculated using the fact that
\begin{align}
    \delta(E - E_{k}) = 
    \frac{\delta(k-k_p)+\delta(k+k_p)}{|E'_{k_p}|},
\end{align}
where the prime denotes differentiation with respect to $k$, which is performed numerically.

In quasi-2D the approach is similar, but slightly more involved. In this case, for integrating over $\kv$, we use polar coordinates $\kv=(k,\theta)$. We begin by using $N_\theta$-point Gauss-Legendre quadrature to determine nodes $\theta_i$ and weights $w^{\theta}_i$ such that
\begin{align}
    \int  \frac{k \, dk\, d\theta}{E-E_{(k,\theta)}} \approx \sum_i w^\theta_i \int  \frac{k\, dk }{E-E_{(k,\theta_i)}}.
\end{align}
Here, we can effectively repeat the process outlined for quasi-1D, for each value of radial momentum $\theta_i$. That is, for fixed energy $E$ and radial momentum $\theta_i$, we can find the pole positions by solving $E-E_{(k,\theta_i)}$ for $k$, which yields one solution that we denote by $k=k_p$. We then find the principal value using the same method as given above (noting the change in the integration measure of $dk \to k\, dk$). Similarly, the Dirac delta contribution is given by
\begin{align}
    \delta(E-E_{(k,\theta_i)}) = \frac{\delta(k-k_p)}{|E'_{(k,\theta_i)}|},
\end{align}
where the numerical derivative is along $k$.

\section{Determination of effective mass} \label{Appendix:DetEffectiveMass}

As presented in the main text, we find that the stability of the numerical scheme for calculating the Hubbard $U$ term is aided by determining the ratio of effective masses (i.e., $m^H_{\rm eff}/m_{\rm eff}$) from the imaginary part of the $T$ matrices (see Eq.~\eqref{Eq:EffMassDetermine}). We now show that this method of calculating the effective mass ratio is equivalent to directly deriving the effective masses from a long-wavelength expansion of the energy dispersion (see Eq.~\eqref{Eq:EffectiveMassLongWavelength}). In particular, we demonstrate that the imaginary part of the inverse of the $T$ matrix is proportional to the effective mass, with a constant of proportionality that is identical for both the Hubbard and the exact $T$ matrix.

To begin, we consider the imaginary component of the Hubbard $T$ matrix, which can be simplified as follows:
\begin{align}
    \Im[T^{-1}_H(\pv \to \zerov)] &= \pi \sum_{\kv \in \rm{1BZ}} \delta(\epsilon_{\pv}^H-\epsilon_{\kv}^H)\\
    &=
        \begin{cases}
            \frac{m^H_{\rm eff}}{p}, & \text{for 1D}, \\[0.8em]
            \frac{m^H_{\rm eff}}{2 }, & \text{for 2D}.            
        \end{cases}
\end{align}
This reveals a direct proportionality between the imaginary component of the inverse Hubbard $T$ matrix and the Hubbard effective mass. The analysis of the exact $T$ matrix is more intricate but follows a similar path. To illustrate this, we introduce a normalized dimer ket
\begin{align}
    \ket{\alpha} = \frac{1}{\sqrt{\sum_{\nuv} \mathcal{H}^{\zerov \nuv}_{\pv \zerov \zerov }{}^2}} \sum_{\nuv} \mathcal{H}^{\zerov \nuv}_{\pv \zerov \zerov } \ket{\nuv},
\end{align}
and apply Gram-Schmidt orthogonalization to generate linearly independent normalized kets $\{\ket{\beta_i}\}$, forming a complete basis for the dimer with $\{\ket \alpha\} \cup \{\ket{ \beta_i}\}$. Consequently, the $T$ matrix is expressed as
\begin{align}
    T(\pv) = g^2 |\phi_{0}|^2 \left(\sum_{\nuv} \mathcal{H}^{\zerov \nuv}_{\pv \zerov \zerov }{}^2 \right) \bra{\alpha} \hat D(\zerov, E_{\pv}) \ket{\alpha}.
\end{align}
Moreover, the sole imaginary contribution to the polarization bubble resides within the ${ \ket{\alpha} }$ component. To understand this, consider
\begin{align}
    \Im[\Pi^{\zerov}_{\nuv \nuv'}] &= - \pi |\phi_{0}|^2 \sum_{\kv \in \rm{1BZ}} \mathcal{H}^{\zerov \nuv}_{\kv \zerov \zerov} \delta(E_{\pv} - E_{\kv}) \mathcal{H}^{\zerov \nuv'}_{\kv \zerov \zerov} \nonumber\\
    &=
    \begin{cases}
        -  |\phi_{0}|^2 \mathcal{H}^{0 \nu}_{p 0 0} \mathcal{H}^{0 \nu'}_{p 00} \frac{m_{\rm{eff}}}{|p|}, & \text{for 1D}, \\[0.8em]
        - |\phi_{0}|^2  \mathcal{H}^{\zerov \nuv}_{\pv \zerov \zerov}  \mathcal{H}^{\zerov \nuv'}_{\pv \zerov \zerov} \frac{m_{\rm eff}}{2}, & \text{for 2D},
    \end{cases}
\end{align}
which indicates that $\Im[\hat \Pi^{\zerov}]$ is proportional to $\dyad{\alpha}$. Consequently, we find that
\begin{align}
     \Im &\left[ \left(\bra{\alpha }\hat D(\zerov, E_{\pv}) \ket{\alpha} \right)^{-1} \right]\\
     &{} \quad = \begin{cases}
        g^2 |\phi_{0}|^2 \left(\sum_{\nu} \mathcal{H}^{0 \nu}_{p 0 0}{}^2 \right)  \frac{m_{\rm{eff}}}{|p|}, & \text{for 1D}, \\[0.8em]
        g^2  |\phi_{0}|^2 \left( \sum_{\nuv} \mathcal{H}^{\zerov \nuv}_{\pv \zerov \zerov}{}^2  \right) \frac{m_{\rm eff}}{2}, & \text{for 2D},
     \end{cases}
\end{align}
as seen from blockwise inversion. Thus, the imaginary component of the exact inverse $T$ matrix mirrors the Hubbard case, where
\begin{align}
    \Im \left[ T^{-1}(\pv \to \zerov) \right]    &=
        \begin{cases}
            \frac{m_{\rm eff}}{|p|}, & \text{for 1D}, \\[0.8em]
            \frac{m_{\rm eff}}{2 }, & \text{for 2D}.            
        \end{cases}
\end{align}
This demonstrates that the ratio of the imaginary components of the inverse exact and Hubbard $T$ matrices serves as an equivalent approach to calculating the ratio of effective masses.

\bibliography{references}

\end{document}